\documentclass[aps,prl,preprint,groupedaddress]{revtex4-2}
\usepackage{graphicx}
\usepackage{epstopdf}
\usepackage{color,soul}
\usepackage{hyperref}
\usepackage{lineno}
\usepackage{amsmath}
\usepackage{mathtools}
\usepackage[normalem]{ulem}
\begin{document}

\title{Giant magnetoresistance, Fermi surface topology, Shoenberg effect and vanishing quantum oscillations in type-II Dirac semimetal candidates MoSi$_2$ and WSi$_2$}
\author{Orest Pavlosiuk$^{1,*}$}
\author{Przemys{\l}aw Wojciech Swatek$^2$}
\author{Jian-Ping Wang$^{2}$}
\author{Piotr Wi{\'{s}}niewski$^1$}
\author{Dariusz Kaczorowski$^{1,3}$} 
\affiliation{$^1$~Institute of Low Temperature and Structure Research, Polish Academy of Sciences, Ok\'{o}lna 2, 50-422 Wroc{\l}aw, Poland}
\affiliation{$^2$~Electrical and Computer Engineering Department, University of Minnesota, Minneapolis, Minnesota 55455, USA}
\affiliation{$^3$~Institute of Molecular Physics, Polish Academy of Sciences, Smoluchowskiego 17, 60-179 Pozna{\'{n}}, Poland}

\affiliation{$^*$~Corresponding author: o.pavlosiuk@intibs.pl}
%
\begin{abstract}
We performed comprehensive theoretical and experimental studies of the electronic structure and the Fermi surface topology of two novel quantum materials, MoSi$_2$ and WSi$_2$. 
The theoretical predictions of the electronic structure in the vicinity of the Fermi level was verified experimentally by thorough analysis of the observed quantum oscillations in both electrical resistivity and magnetostriction. 
We established that the Fermi surface sheets in MoSi$_2$ and WSi$_2$ consist of 3D dumbbell-shaped hole-like pockets and rosette-shaped electron-like pockets, with nearly equal volumes. 
Based on this finding, both materials were characterized as almost perfectly compensated semimetals. 
In conjunction, the magnetoresistance attains giant values of $10^4$ and $10^5\,\%$ for WSi$_2$ and MoSi$_2$, respectively. 
In turn, the anisotropic magnetoresistance achieves $-95$ and $-98\,\%$ at $T=2$\,K and in $B=14$\,T for WSi$_2$ and MoSi$_2$, respectively.    
Furthermore, for both compounds we observed the Shoenberg effect in their Shubnikov-de Haas oscillations that persisted at as high temperature as $T=25$\,K in MoSi$_2$ and $T=12$\,K in WSi$_2$. 
In addition, we found for MoSi$_2$ a rarely observed spin-zero phenomenon. 
Remarkably, the electronic structure calculations revealed type-II Dirac cones located near 480\,meV and 710\,meV above the Fermi level in MoSi$_2$ and WSi$_2$, respectively. 
\end{abstract}
\maketitle

\section*{Introduction}

Topological semimetals (TSMs) constitute the most numerous and diverse group of topological materials.\cite{Armitage2018,Hu2019a,Lv2021} 
They form a special subclass of gapless electronic phases that exhibit topologically stable crossings in the energy dispersion of their bulk electronic states.	
Different types of TSMs can be distinguished into 3 main groups based on discriminating characteristics, i.e., ($i$) type of low-energy excitation (Dirac and Weyl semimetals), ($ii$) codimension, e.g., type of band crossing (point-like, nodal-line and multi-fold band crossings), and ($iii$) preserving of Lorentz invariance (type-I and type-II topological semimetals).\cite{Gao2019b} 
In nodal-line semimetals, the gaps close along lines or loops in the Brillouin zone (BZ) rather than at isolated points as in type-I Dirac/Weyl semimetals.\cite{Armitage2018,Hirayama2017} 
In type-II TSMs, the Dirac/Weyl cone exhibits strong tilting so that the characteristic crossing point appears as a contact point between an electron pocket and a hole pocket of the Fermi surface sheets. 
This behavior is a straightforward consequence of violation of the Lorentz symmetry in the crystal structure.\cite{Soluyanov2015} 
Depending on the position of the chemical potential, these topologically non-trivial electronic features can be observed in a form of peculiar electron transport properties associated with enhanced carrier mobility and chiral magnetic anomaly, both being a consequence of non-zero Berry curvature in the momentum space.\cite{Hu2019a} 

Besides the unprecedented importance for fundamental science, TSMs also offer an intriguing and promising opportunity for device design, revolutionizing future spin-orbital torque based low-power memory and computational capabilities, quantum computing hardware, as well as laser technology.\cite{deLeon2021,Wang2020,Shao2021} 
Because the electronic transport properties play an important role in device modeling and effectiveness, it is crucial to develop accurate and detailed predictions of electrical conductivity and magnetoconductivity, and more fundamentally, electron band dispersion in the vicinity of the Fermi level.
Additionally, semiconductor industry imposes additional requirements on usable materials, requiring them to be cheap, stable in different environmental conditions, non-toxic, and easy-obtainable by large-scale industrial methods. 

MoSi$_2$ and WSi$_2$ meet all the above requirements, including a compatible growth process as thin films.\cite{Krontiras1987,Afzal2021,Mohammed2020}
Recent works on MoSi$_2$ and WSi$_2$ demonstrated that both compounds may exhibit some topological features, such as non-trivial Berry phase (extracted from the quantum oscillations of magnetization) and extremely large magnetoresistance (XMR).\cite{Matin2018,Mondal2020} 
While the first effect is usually interpreted as the presence of non-trivial electronic states near the Fermi level, several other mechanisms (related to both topologically trivial and non-trivial character of electronic structure) have been suggested as explanations for XMR in various materials.\cite{Shekhar2015c,Pavlosiuk2016f,He2016a,Tafti2016a,Li2016f,Liang2014} 
Nevertheless, in the preliminary works on MoSi$_2$ and WSi$_2$, it has been emphasized that additional experimental and theoretical studies on high-quality materials are desired to conclude on possible topologically non-trivial electronic properties.\cite{Matin2018,Mondal2020} 
There is also an additional controversy in the interpretation of the results of the quantum oscillations analyses performed for MoSi$_2$ in Refs.~\onlinecite{Matin2018} and \onlinecite{Ruitenbeek1987}. Two independent research groups ascribed the same frequencies of quantum oscillations to different extreme cross-sections of the Fermi pockets, causing divergent statements about the shape of the Fermi surfaces.  

Remarkably, MoSi$_2$ and WSi$_2$ crystallize with the $I4/mmm$ space group that is adopted also by the $M\!A_3$ ($M$=V, Nb, Ta; $A$=Al, Ga, In) compounds, which are archetypal type-II Dirac semimetals.\cite{Chang2017a} 
The existence of tilted Dirac cones can lead to some unique physical properties like Klein tunneling\cite{OBrien2016} and anomalous Hall effect,\cite{Zyuzina2016} which can be utilized in novel ultrasensitive magnetic sensors and memories.\cite{Wang2021,Ohuchi2018} 
However, there are many challenges and critical issues in growing single-crystalline $M\!A_3$ by deposition techniques due to high vapor pressure of Al, Ga and In.\cite{Harsha2005} In contrast, it has been shown that thin films of MoSi$_2$ can be easily prepared on different substrates,\cite{Afzal2021,Mohammed2020} thus providing an ideal platform for both basic and applied research.

In order to verify whether MoSi$_2$ and WSi$_2$ can be classified as topological semimetals, and thus can be considered as useful material for modern technology applications, we performed detailed calculations of their electronic structure, carried out meticulous investigations of their electron transport properties, and analyzed in details the observed quantum oscillations of electrical resistivity and magnetostriction.  
We obtained very good agreement between theoretical and experimental datasets. 
The calculated and experimentally obtained Fermi surface sheets of both materials are almost identical. 
In addition to the important information about the Fermi surface topology, we found that quantum oscillations of electrical resistivity in MoSi$_2$ and WSi$_2$ demonstrate pronounced magnetic interaction effect, which is also called the Shoenberg effect. 
In both compounds this effect appears as the combination frequencies in the fast Fourier transform spectra of quantum oscillations.
In MoSi$_2$ this effect was noticed at a record-high temperature of 25\,K. 
Furthermore, the quantum oscillations of electrical resistivity in MoSi$_2$ show the spin-zero effect, the vanishing of the fundamental frequency of quantum oscillations due to the fact that for the certain directions of magnetic field application, the spin factor (in the Lifshitz-Kosevich theory) becomes zero.

In contrast to the relatively common observation of quantum oscillations in electrical resistance and magnetization, quantum oscillations of magnetostriction are rarely reported. 
In our work, we observed these oscillations in MoSi$_2$ and confirmed that they can be a comprehensive technique for mapping Fermi surfaces in semimetals. 

\section*{Methods}

Single crystals of MoSi$_2$ and WSi$_2$ were grown by the Czochralski technique. 
The synthesis protocol consists of two steps and is as follows. 
First, polycrystalline precursors were synthesized by arc-melting of stoichiometric amounts (\{Mo,W\}:Si = 1:2) of the elemental constituents with chemical purities Mo (99.97 wt.\%), W (99.95 wt.\%) and Si (99.9999 wt.\%). 
Next, those polycrystalline samples were used to grow single crystals using a tetra arc-furnace, the syntheses were carried out under argon atmosphere.
The obtained single crystals were studied by X-ray Laue backscattering with a Proto LAUE COS system, in order to check their quality and orient them along special crystallographic direction.

The crystal structure of powdered MoSi$_2$ and WSi$_2$ single crystals was confirmed by powder X-ray diffraction, using PANanalytical X’pert Pro diffractometer with Cu-K$\alpha$ radiation. 
The obtained X-ray diffractograms were analyzed with FullProf software (Rietveld method was used).\cite{Rodriguez-Carvajal1993}

The electrical transport measurements were carried out using a standard four-probes technique with a Quantum Design Physical Property Measurement System (PPMS) equipped with a horizontal rotator. 
Rectangular-shaped samples were cut from the oriented single crystals by wire saw, electrical contacts were made of 50-$\rm\mu$m silver wires which were attached to the sample by silver epoxy paste. 
A miniaturized capacitance dilatometer\cite{Kuchler2012} and the PPMS platform were used for the magnetostriction measurements.   

Electronic structure calculations were performed with the all-electron general potential linearized augmented plane-wave method as implemented in the Elk code.\cite{ELKwww,Mller2020}
The exchange and correlation effects were treated using GGA in the form proposed by Perdew, Wang and Ernzerhof.\cite{Perdew1996} 
The spin-orbit coupling (SOC) was included as a second variational step, using scalar-relativistic eigenfunctions as the basis, after the initial calculation was converged to self-consistency. 
The Monkhorst–Pack special $k$-point scheme with
22 x 22 x 17 mesh was used in the first Brillouin zone sampling, and the muffin tin radius ($RK_{max}$) was set to 8.\cite{Monkhorst1976,9780387296845} 
For the Fermi surface, the irreducible Brillouin zone was sampled by 15620 $k$-points to ensure accurate determination of the Fermi level.\cite{Kokalj2003} 
Quantum oscillations frequencies were calculated using the Supercell K-space Extremal Area Finder tool.\cite{Rourke2012}
For the reference data of DOS similar calculations were performed with a full potential all-electron local orbital code FPLO-14.00-49,\cite{Koepernik1999,Ylvisaker2009,Lejaeghere2016} using the same type of the exchange-correlation potential as above. 
In all calculations, the experimental lattice parameters of MoSi$_2$ and WSi$_2$ obtained here were assumed.

\section*{Results and discussion}

\subsection*{Crystal structure and electronic structure calculations}

\begin{figure}[h]
	\includegraphics[width=16cm]{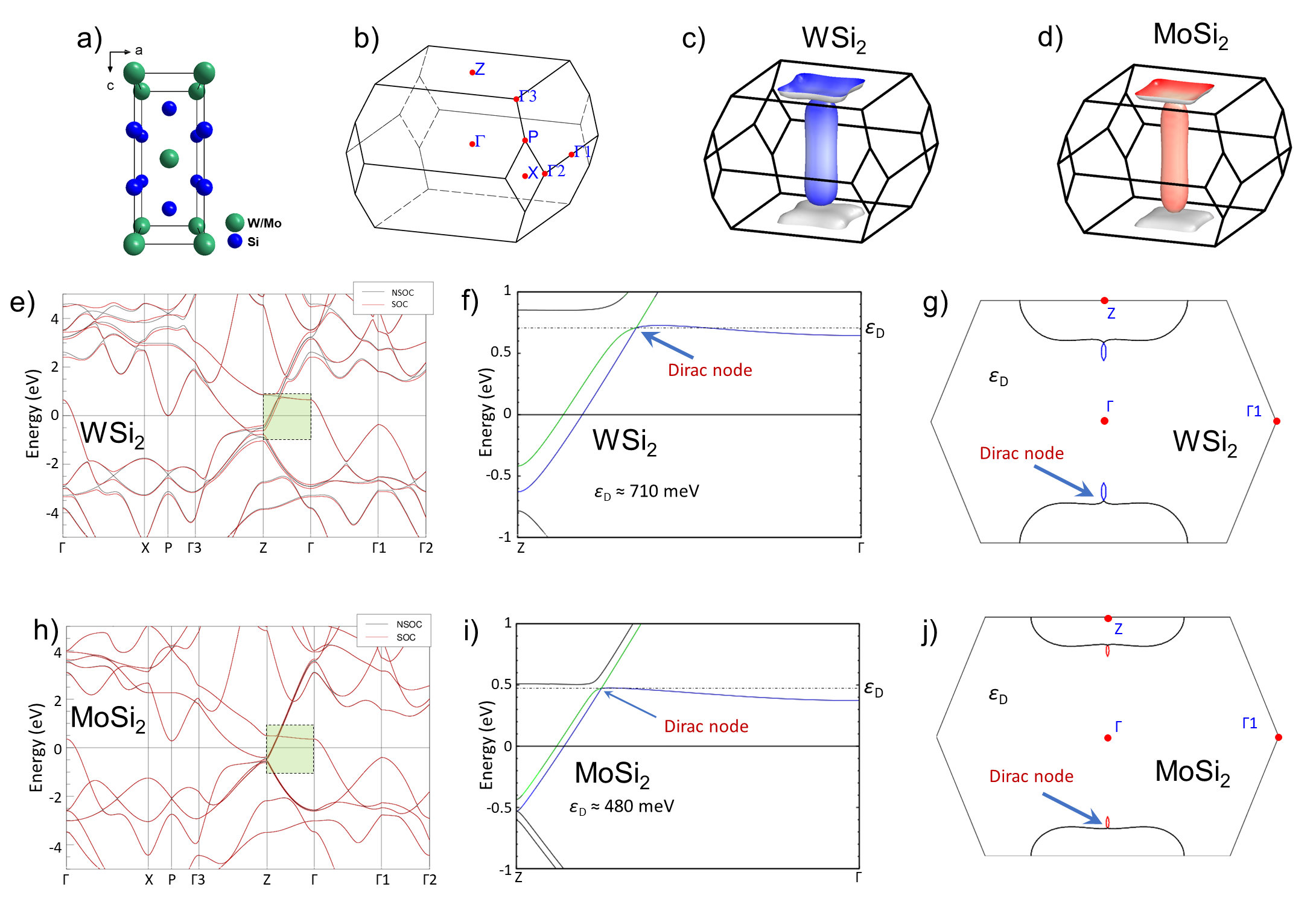}
	\caption{(a) Unit cell of \{W,Mo\}Si$_2$, green spheres correspond to W or Mo atoms, blue spheres denote Si atoms. (b) Brillouin zone (with high-symmetry points) of \{W,Mo\}Si$_2$ for the crystal structure shown in Fig.~\ref{band_structure}a. (c,d) Fermi surfaces of WSi$_2$ an MoSi$_2$. (e,i) Electronic band structures of WSi$_2$ and MoSi$_2$, green rectangles highlight the area in the vicinity of type-II Dirac crossings. Black lines correspond to band structure calculated without SOC (NSOC) and red lines correspond to band structure calculated with SOC. (f,j) Zoomed in electronic structures (with SOC) near the Dirac node indicated by green rectangle in (e,i). Constant energy contour at the energy of Dirac node (g,j).        
		\label{band_structure}}
\end{figure}

X-ray powder diffraction analysis confirmed that both compounds crystallize with the tetragonal crystal structure of the space group $I{\rm4}/mmm$ (see Fig.~\ref{band_structure}a). 
The crystal structures are layered and they are formed by stacking of \{Mo,W\}-Si-\{Mo,W\}-Si-\{Mo,W\} slabs along the $c$-axis.
The obtained values of lattice parameters ($a=3.21395$\,\r{A} and $c=7.83128$\,\r{A} for WSi$_2$; $a=3.20519$\,\r{A} and $c=7.84461$\,\r{A} for MoSi$_2$) as well as atomic coordinate of Si atom ($z=0.33247$ and $z=0.33497$ for WSi$_2$ and MoSi$_2$, respectively) are very close to those reported in the literature.\cite{Tanaka2000} 
To provide better agreement with our experimental results, we used these crystal structure parameters during our theoretical calculations, results of which are shown below (see Fig.~\ref{band_structure}). 

The Fermi surfaces of MoSi$_2$ and WSi$_2$ have been studied by means of theoretical calculations,\cite{Bhattacharyya1985,Shugani2015,Ruitenbeek1987,Matin2018,Mondal2020} and quantum oscillations analysis.\cite{Ruitenbeek1987,Matin2018,Mondal2020}
All these studies have shown that both compounds are semimetals with Fermi surfaces containing one electron-like and one hole-like sheet. 
Interestingly, in two recent experimental papers,\cite{Matin2018,Mondal2020} the authors discussed the possibility of the existence of non-trivial topological states in both compounds. 
However, these non-trivial states have not been tackled by their theoretical calculations. 
Therefore, we performed detailed calculations of electronic structure of both materials with the purpose of looking for topologically non-trivial states.

The bulk Brillouin zone with marked high-symmetry points is shown in Fig.~\ref{band_structure}b, with the electronic structures of MoSi$_2$ and WSi$_2$ shown in Fig.~\ref{band_structure}e,h. 
The impact of spin-orbit coupling is shown by comparing the results without SOC (black solid lines) and with SOC (red solid lines).
There are several gapless nodes at high-symmetry $k$-points around the Fermi level, e.g., in the $\Gamma$-$K$ plane, when the SOC is not included.
As the SOC is induced, those Dirac points become gapped and thus pronounced contribution of a spin Berry curvature can be expected. 
The spin Berry curvature seems to be inversely proportional to the gap size, in full agreement with theoretical predictions.\cite{Qiao2018}
The only noticeable crossing, robust against SOC, is along the $\Gamma$-$Z$ direction at $\varepsilon_D=480$\,meV and $\varepsilon_D=710$\,meV above the Fermi level for MoSi$_2$ and WSi$_2$, respectively (see Fig.~\ref{band_structure}e,f,h,i).

Both bulk Dirac cones of MoSi$_2$ and WSi$_2$ are formed by two W/Mo valence bands with mainly $d_{xy}$ and $d_{xz+yz}$ orbital characters. 
As each electronic band is doubly degenerate, the isolated local symmetry-protected bands create four-fold degenerate Dirac points. 
Group-theory analysis shows that these two bands belong to different irreducible representations, which are associated with $D_{4h}$ point symmetry. 
Similar band crossing, along the same direction in the BZ, appears in several isostructural materials from the $MA_3$-family (where $M$ = V, Nb, Ta; $A$ = Al, Ga, In) of type-II Dirac semimetals.\cite{Chang2017a}
Type-II Dirac cones in PtSe$_2$ and PtTe$_2$ are formed mostly by Se/Te-$p$ orbitals,\cite{Zhang2017e,Yan2017a} in turn in MoSi$_2$ and WSi$_2$ the Dirac cones are formed mainly by the $d$ orbitals. 

The Fermi surfaces of WSi$_2$ and MoSi$_2$ are shown in Fig.~\ref{band_structure}c,d. 
Both compounds have similar Fermi surfaces, comprising two pockets. 
A hole-like pocket is centered at $\Gamma$ point and electron-like pocket is centered at $Z$ point. 
The coexistence of the electron- and hole-like carriers is consistent with the Hall resistivity results (see Supplementary Information). 
To add clarity, the constant energy contours at $\varepsilon_n(k)=\varepsilon_D$ are shown in Fig.~\ref{band_structure}g,j. 
For both compounds, the hole-like and the electron-like pockets touch each other at the Dirac node. 

\subsection*{Electrical resistivity and magnetoresistance}

\begin{figure}[h]
	\includegraphics[width=8cm]{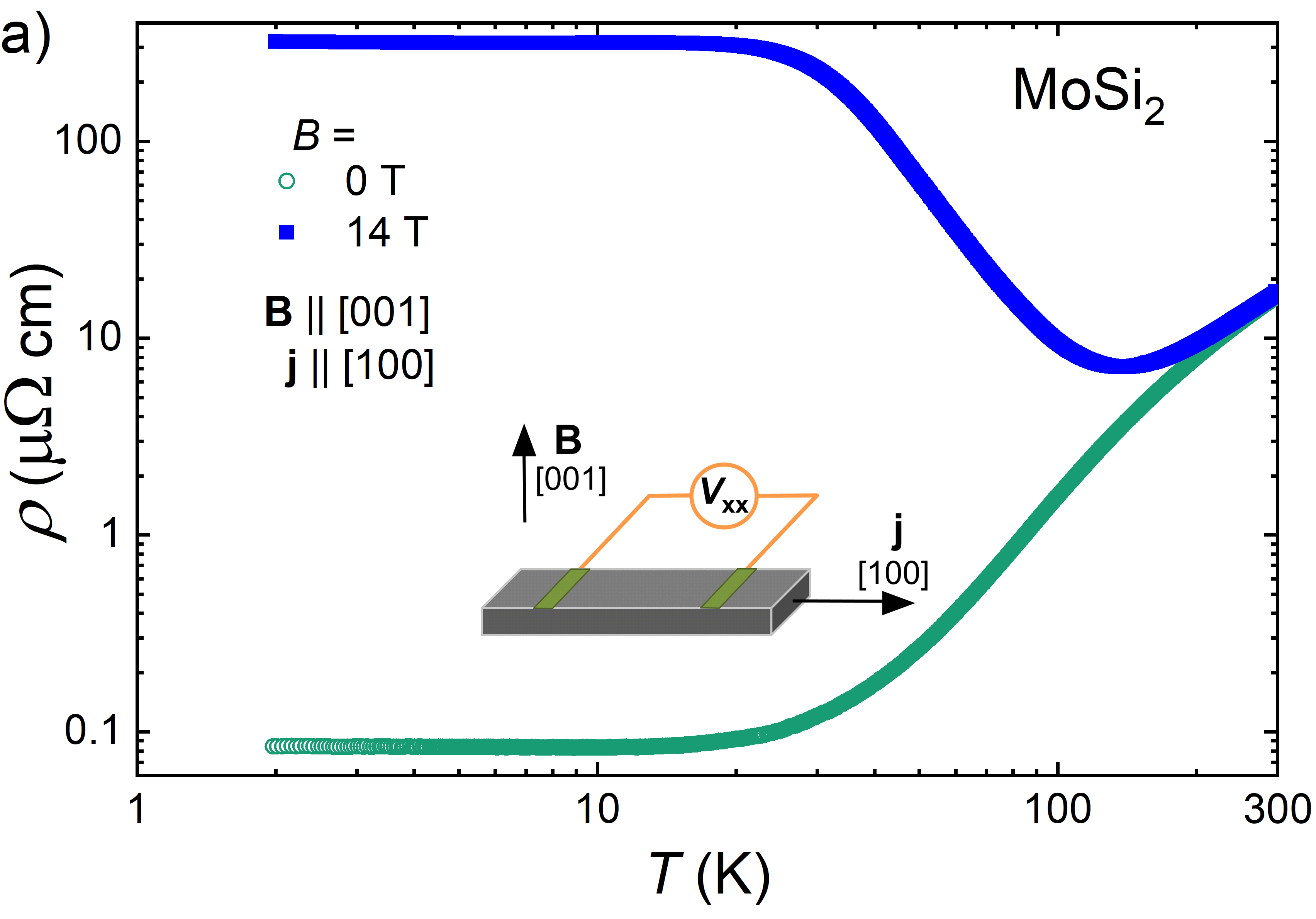}
	\includegraphics[width=8cm]{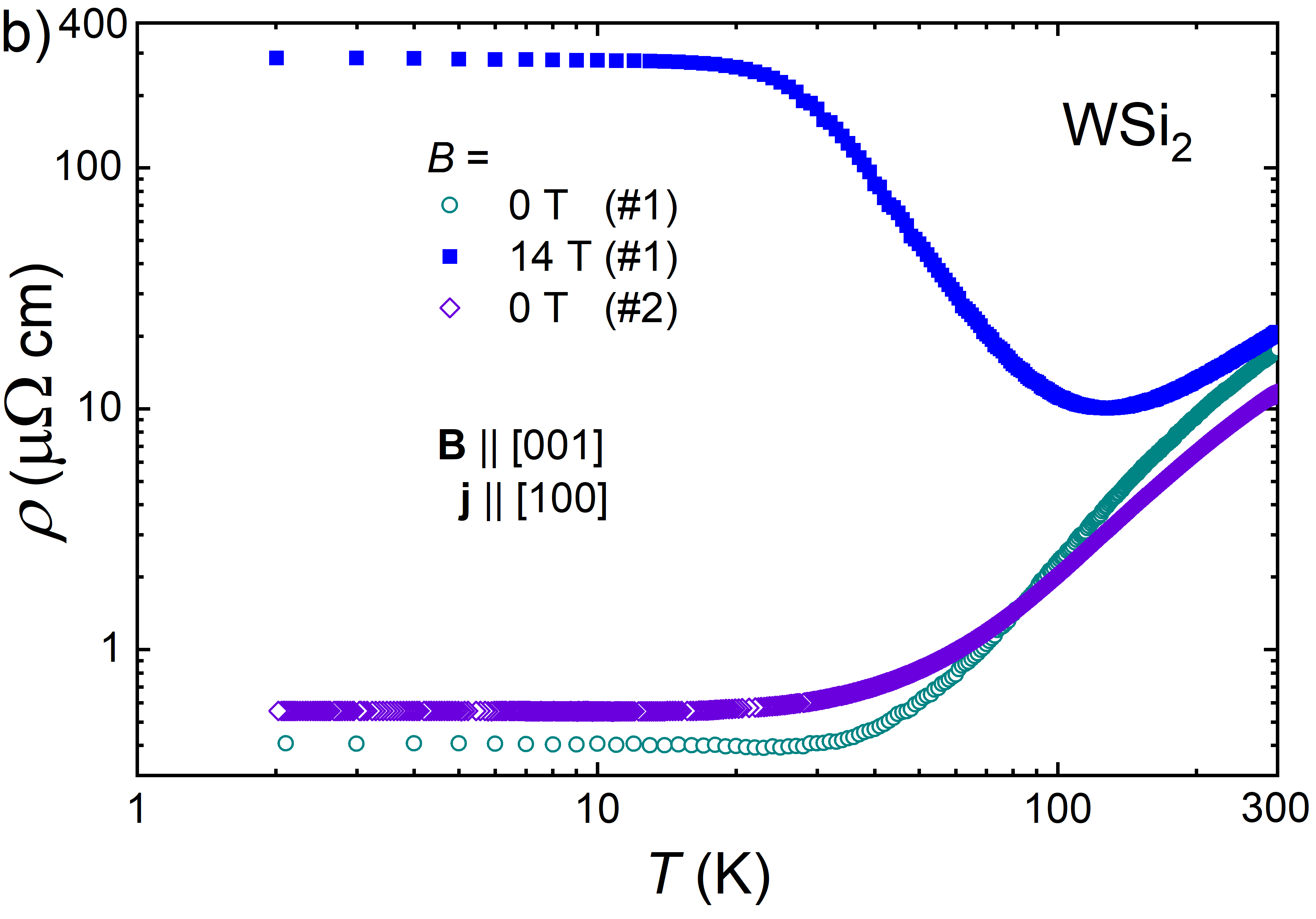}
	\includegraphics[width=8cm]{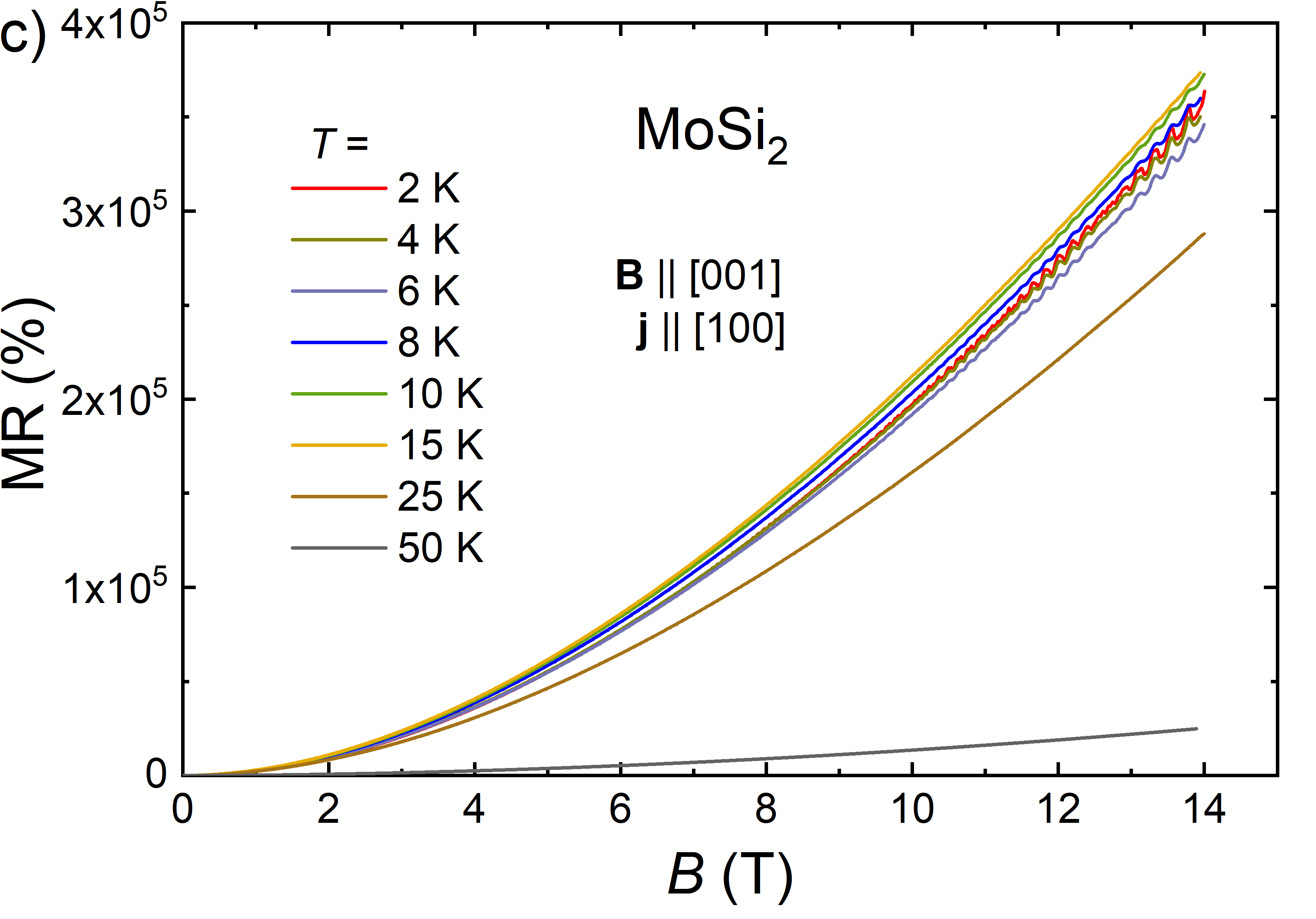}
	\includegraphics[width=8cm]{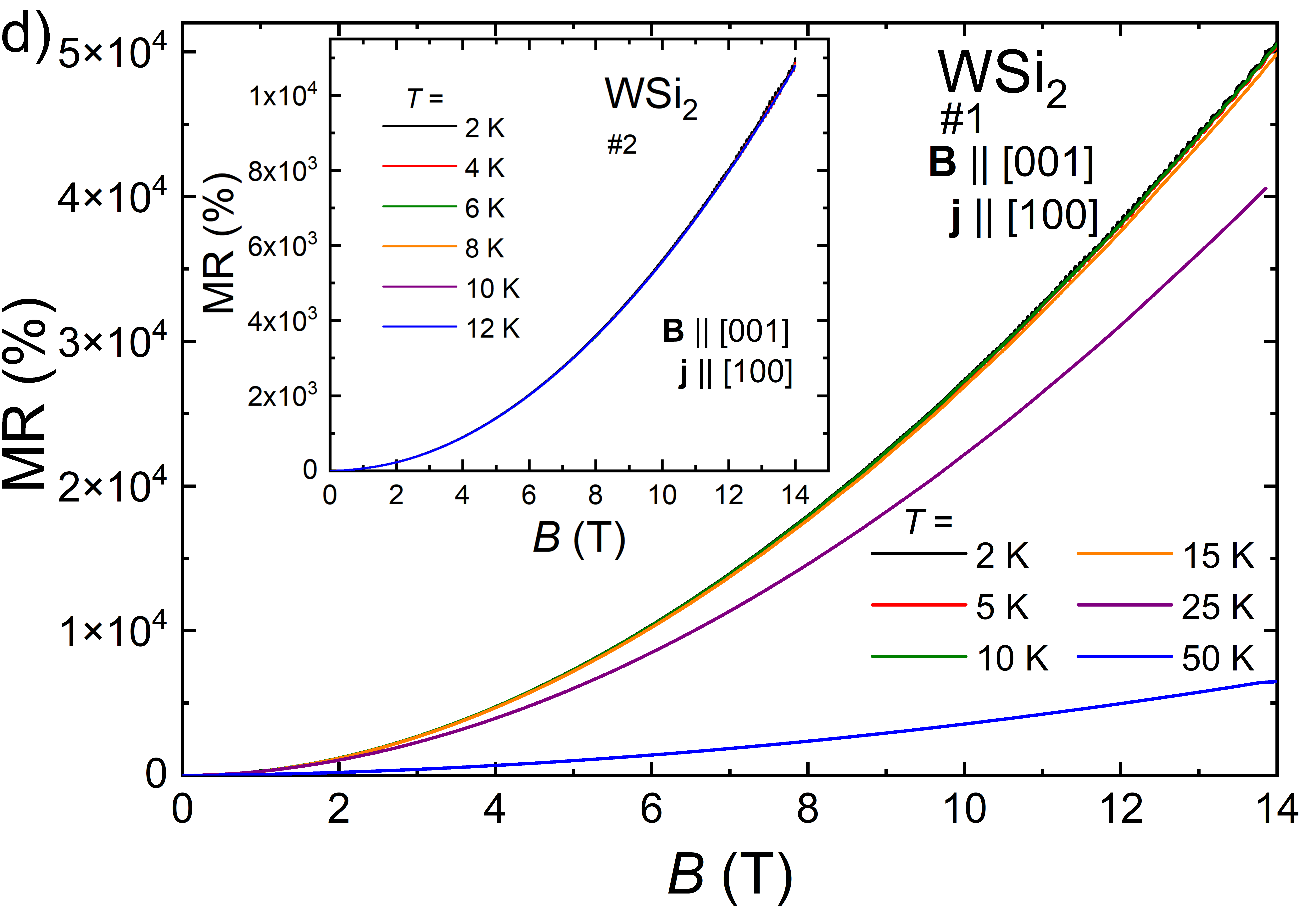}
	\caption{Electrical resistivity as a function of temperature in a log-log scale, measured in zero magnetic field and in applied magnetic field of 14\,T for MoSi$_2$ (a) and WSi$_2$ (sample $\#1$ and sample $\#2$) (b). Magnetoresistance as a function of magnetic field, measured at several temperatures for MoSi$_2$ (c) and sample $\#1$ of WSi$_2$ (d). Inset to (d) shows magnetoresistance isotherms obtained for sample $\#2$ of WSi$_2$. The measurement geometry is shown in the inset to (a).
		\label{rho(T)}}
\end{figure}

As the results of our theoretical calculations point to the possible formation of type-II Dirac cones in MoSi$_2$ and WSi$_2$, the next stage of our research was experimental verification of the calculated electronic structure. 
It should be noted that according to our calculations, the topological non-trivial states are located somewhat above the Fermi level, which means that their direct contribution to the overall electronic transport properties can be not visible.
Nevertheless, if this calculated electronic structure can be verified experimentally, even in the vicinity of the Fermi level only, it will be reasonable to undertake further research to shift the Fermi level closer to the type-II Dirac points. 
There are several strategies to do this. 
For example, the Fermi level tuning in heterostructures can be achieved by doping, defect control, epitaxial thin film growth on different substrates, and a recently proposed mechanism based on the cooperative effect of charge density waves and non-symmorphic symmetry.\cite{Aitani2013,Lei2021}

To verify experimentally the calculated electronic structure near the Fermi level, we focused on quantum oscillations of electrical resistivity, because to our knowledge only the quantum oscillations of magnetization for MoSi$_2$ and WSi$_2$ have previously been analyzed.\cite{Ruitenbeek1987,Matin2018,Mondal2020}  
The phenomenon of quantum oscillations underlies a simple but powerful and accurate technique of direct mapping of the Fermi surface.\cite{Shoenberg1984} 
In contrast to the angle-resolved photoemission spectroscopy or scanning tunneling spectroscopy, this technique does not require complex and time-consuming sample preparation combined with sophisticated equipment. 
However, despite simple methodology, high-quality and pure single-crystalline samples with large electron mean free path are required to observe oscillations.

In conjunction with the Laue diffraction data (see Supplementary Information), large residual resistivity ratio ($RRR=\rho(300K)/\rho(2K)$) of our samples ($RRR=193$, $RRR=43$ and $RRR=21$ for MoSi$_2$, WSi$_2$ (sample $\#1$) and WSi$_2$ (sample $\#2$), respectively (see Fig.~\ref{rho(T)})), confirms high quality of our samples, which allows detection of quantum oscillations. 

The $\rho(T)$ dependences recorded in zero magnetic field and in $B=14$\,T for both compounds are presented in Fig.~\ref{rho(T)}a,b. 
In $B=0$\,T, $\rho(T)$ shows metallic-like behavior, however, in $B=14$\,T and below $\sim\!130$\,K, $\rho$ starts to increase with $T$ lowering and saturates at $T<15$\,K.  
This magnetic field-induced resistivity plateau has frequently been observed in topologically trivial\cite{Pavlosiuk2016f,Pavlosiuk2017,Pavlosiuk2018a} and non-trivial\cite{Ali2014,Shekhar2015c,Singha2016a} semimetals.
There are a few possible origins of this behavior associated with metal-insulator transition in topological semimetals,\cite{Li2016f,Singha2016a,Hosen2017} perfect or nearly perfect electron-hole compensation in conventional semimetals,\cite{Wang2015d,Xu2017f} or Lifshitz transition.\cite{Wu2015} 

The huge difference in $\rho$ values obtained in zero and applied magnetic fields at low temperatures confirms that MoSi$_2$ and WSi$_2$ are materials with extremely large magnetoresistance (XMR). 
As is shown in Fig.~\ref{rho(T)}c,d at $T<15$\,K and in $B=14$\,T, magnetoresistance (${\rm{MR}}=\rho(B)/\rho(0)-1$) achieves values on the order of $10^5\%$ and $10^4\%$ for MoSi$_2$ and WSi$_2$ (samples \#1 and \#2), respectively, which resemble those reported in Refs.~\onlinecite{Matin2018,Mondal2020}. 
In the former work, the authors attributed XMR in MoSi$_2$ to the Fermi surface reconstruction due to the Zeeman effect. 
As it was suggested earlier, in WSi$_2$, ultrahigh mobilities of near-perfectly balanced electron- and hole-like carriers lead to XMR.\cite{Mondal2020} 
There are also several other mechanisms explaining XMR in various materials: (i) moderate carrier compensation with substantial mobility difference\cite{He2016a} (ii) $d\!-\!p$ orbital mixing combined with carrier compensation\cite{Tafti2016a} (iii) magnetic field induced metal-insulator-like transition\cite{Li2016f} (iv) topological protection from  backscattering.\cite{Liang2014}  

The values of MR observed for WSi$_2$ and MoSi$_2$ are of the same order of magnitude as those reported for several other topological semimetals.\cite{Ali2014,Shekhar2015c,Singha2016a}
Based on the results of our electronic structure calculations, quantum oscillations analysis (see below) and Hall effect data (see Supplementary Information), it can be concluded that nearly perfect compensation of carriers and their high mobility play the dominant role in the magnetotransport properties of both studied materials.

\subsection*{Shubnikov-de Haas effect}

\begin{figure}[h]
	\includegraphics[width=8cm]{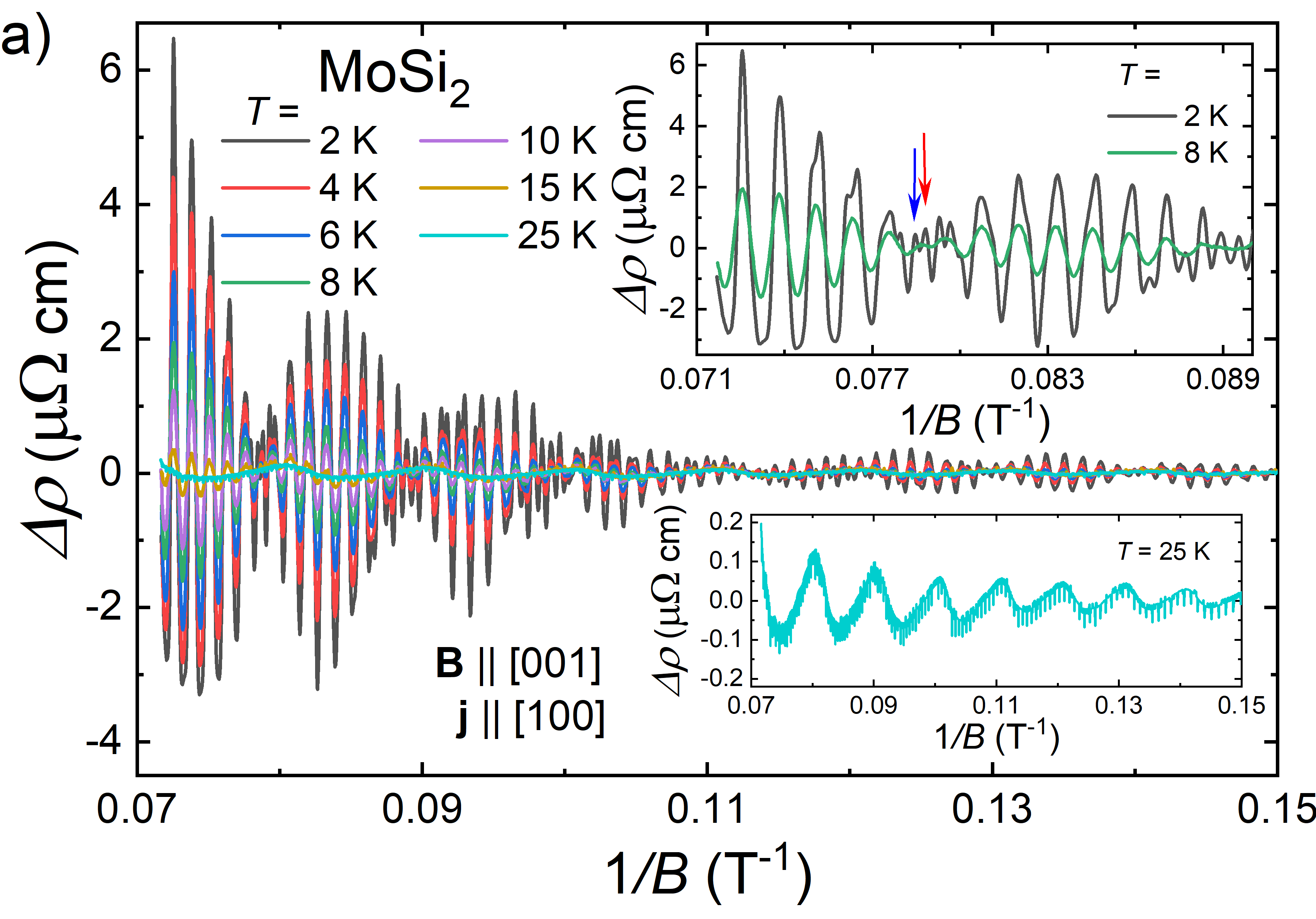}
	\includegraphics[width=8cm]{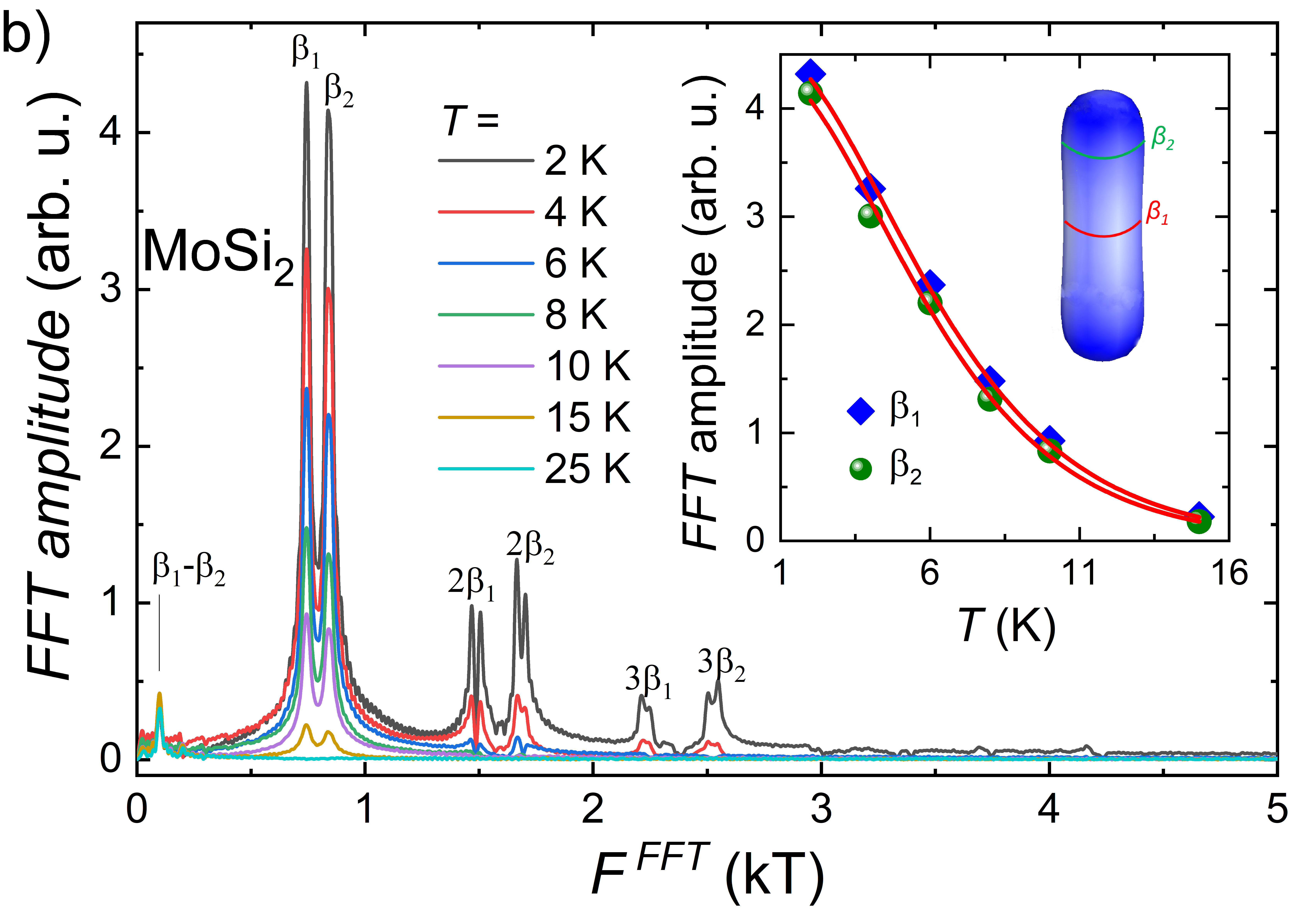}
	\includegraphics[width=8cm]{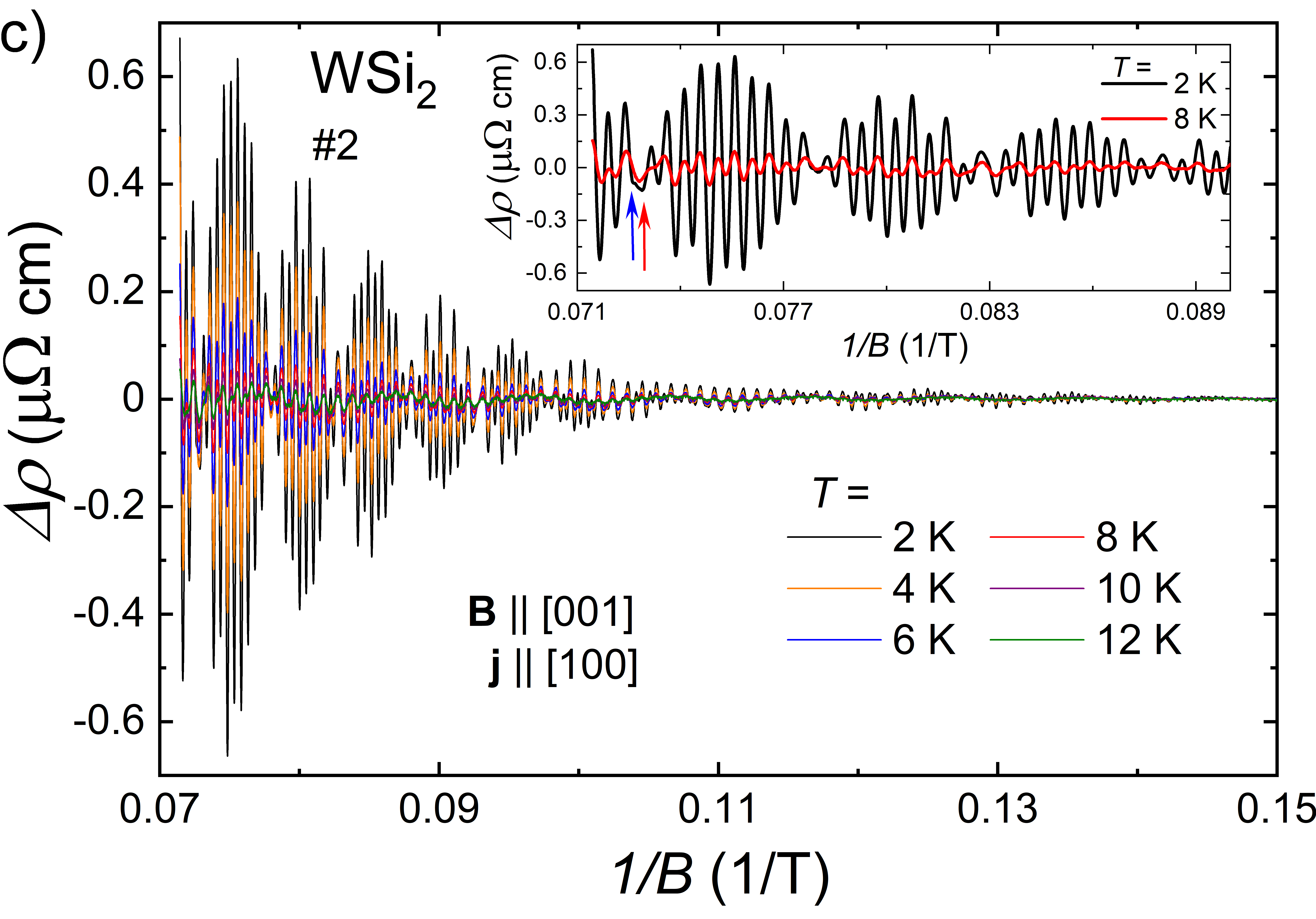}
	\includegraphics[width=8cm]{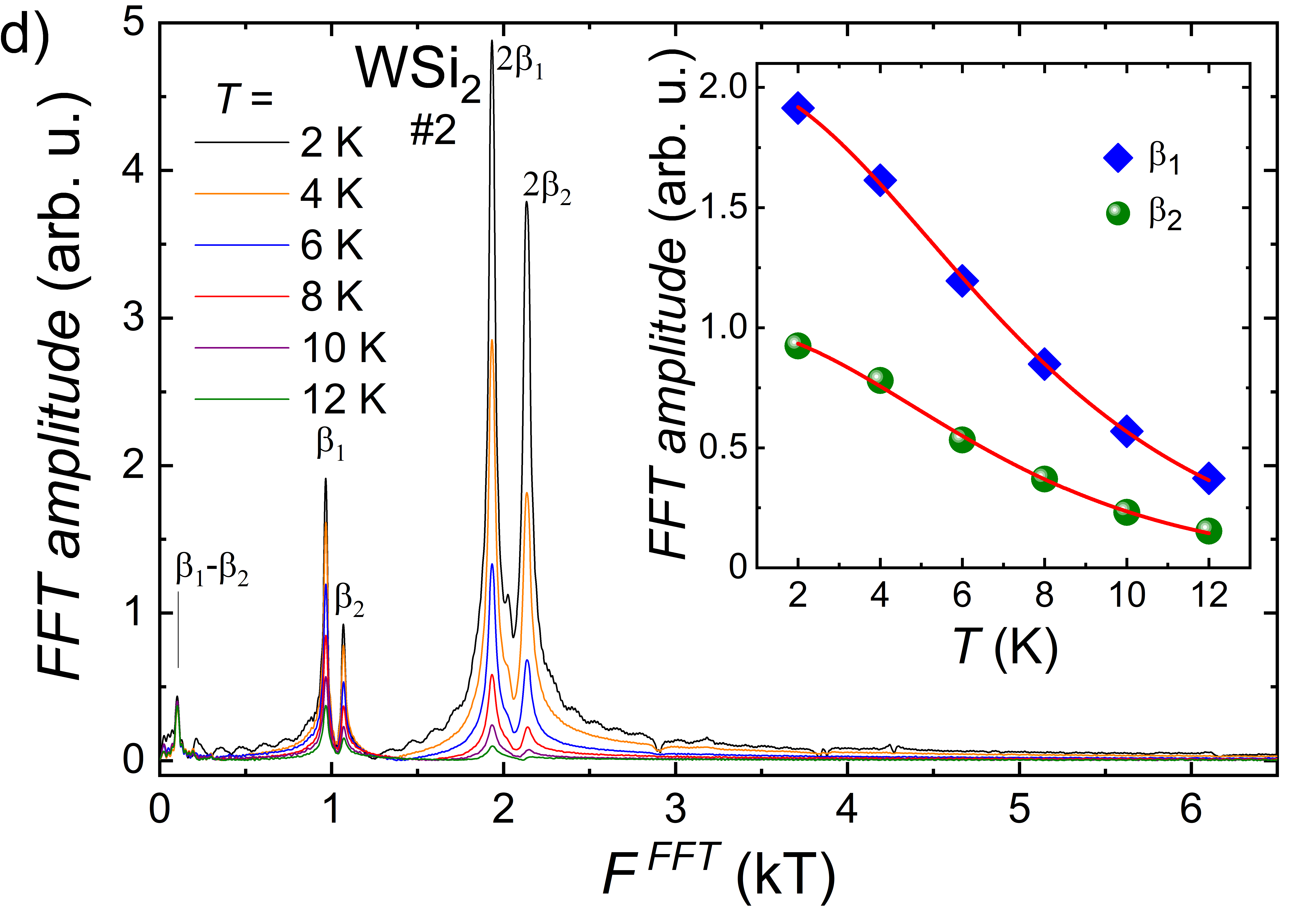}
	\caption{Oscillating part of electrical resistivity as a function of inverted magnetic field at several different temperatures for MoSi$_2$ (a) and WSi$_2$ (sample $\#2$) (c). Upper insets to (a) and (c) are the closeups of the data below $0.09$\,T$^{-1}$ at $T=2$ and 8\,K. Blue and red arrows indicate the splitting of peak (a) and valley (c). Lower inset to (a) shows the Shubnikov-de Haas oscillations with the saw-tooth wave shape at $T=25$\,K. Fast Fourier transform spectra for quantum oscillations in MoSi$_2$ (b) and in WSi$_2$ (sample $\#2$) (d). Insets to (b) and (d) show the temperature dependence of the FFT peak height. Solid red lines correspond to fits to $R_{T,i}(T)$ temperature damping factor of the Lifshitz-Kosevich formula (Eq.~\ref{full_LK_eq}). The hole-like Fermi pocket $\beta$ with the extreme cross sections perpendicular to the [001] direction is shown in the inset of (b). 
		\label{SdH_analysis}}
\end{figure}

In order to probe the Fermi surface structure of studied compounds, we thoroughly analyzed SdH quantum oscillations, which were clearly resolved at temperatures up to at least 25\,K and 12\,K for MoSi$_2$ and WSi$_2$, respectively. 
We observed pronounced SdH oscillations in both studied samples of WSi$_2$ (sample $\#1$ and sample $\#2$), but in the following part of paper, only quantum oscillations in sample $\#2$ are discussed.  
As $\rho(B)$ is a superposition of oscillating and non-oscillating signals (see Supplementary Information), we get rid of the latter one contribution by subtraction of the third-order polynomial function from the experimental data.  
The oscillating part of electrical resistivity, $\Delta\rho$, as a function of inverted magnetic field is presented in Fig.~\ref{SdH_analysis}a and Fig.~\ref{SdH_analysis}c for MoSi$_2$ and WSi$_2$, respectively. 
The overall behavior of $\Delta\rho(1/B)$ directly indicates that the oscillating signal contains several frequencies and hence, the Fermi surfaces of both materials can have rather complex shape. 
The observation of beating patterns could be due to the existence of two frequencies with similar absolute values.    
To decompose the oscillations into their constituent frequencies, we used fast Fourier transform (FFT) analysis. Thus obtained spectra are shown in Fig.~\ref{SdH_analysis}b and Fig.~\ref{SdH_analysis}d for MoSi$_2$ and WSi$_2$, respectively. 
At $T=2$\,K, the analysis yields a rich spectra of frequencies ($F_i$), containing of two fundamental frequencies ($F_{\beta_1}$ and $F_{\beta_2}$), together with their difference ($F_{\beta_2-\beta_1}$) and harmonics ($2F_{\beta_1}$, $2F_{\beta_2}$, $3F_{\beta_1}$ and $3F_{\beta_2}$). 
The third harmonics were clearly observed only for MoSi$_2$. 
Two fundamental frequencies obtained from both the FFT analysis and the calculations from first principles  ($F^{calc}_i$) are very similar for both studied compounds (see Table~\ref{FFT_TABLE}). 
Taking into account the growth method (Czochralski technique), the doping and small dislocations may be brought in and thus the Fermi level may be slightly shifted, which may be the reason for the tiny difference between experimental and theoretical values of oscillations frequencies. 
Similar effects are frequently reported for materials growth with the same technique.\cite{Prokes2017}
Interestingly, in the FFT spectra of MoSi$_2$, which were obtained for $T\leq6$\,K, the peaks related to the harmonic frequencies show the double-peak features, while the peaks related to the fundamental frequencies are not split. The similar character of the FFT spectrum has 
been previously reported for Al$_x$Ga$_{1-x}$N$/$GaN heterostructures and it was described as influence of both zero-field spin-splitting effect and Zeeman spin splitting effect.\cite{Tang2006}

\begin{table}
	\centering
	\begin{tabular}{*{7}{l}} \hline\hline 
		{Compound}&&~~~~~~~~~~{$i$\,=}~~~~~~~~&$\beta_2-\beta_1$~~~~&$\beta_1$&$\beta_2$&$\delta$\\\hline
		MoSi$_2$& $F_i$&(T)&~~99~~~~~~~&743~~~&841~~~&~~~-~~~\\
		& $F^{calc}_i$&(T)&~~~-~~~&823~~~&942~~~&6322~~~\\
		& $n_{i}$&(10$^{20}$cm$^{-3}$)&~~~-~~~~&\multicolumn{2}{c}{8.21~~~~}&~~~-~~~\\
		& $n^{calc}_i$&(10$^{20}$cm$^{-3}$)&~~~-~~~~&\multicolumn{2}{c}{8.18~~~~}&8.11~~~\\
		& $m_i^{*}$&($m_0$)&~~~-&0.25~~~&0.26~~~&~~~-~~~\\
		& $m_i^{*calc}$&($m_0$)&~~~-&0.28~~~&0.31~~~&1.54~~~\\\hline
		WSi$_2$& $F_i$&(T)&~~103~~~~~~~&966~~~&1068~~~&~~~-~~~\\
		& $F^{calc}_i$&(T)&~~~-~~~&1111~~~&1239~~~&7046~~~\\
		& $n_{i}$&(10$^{20}$cm$^{-3}$)&~~~-~~~~&\multicolumn{2}{c}{9.75~~~~}&~~~-~~~\\
		& $n^{calc}_i$&(10$^{20}$cm$^{-3}$)&~~~-~~~~&\multicolumn{2}{c}{10.87~~~~}&10.81~~~\\
		& $m_i^{*}$&($m_0$)&~~~-&0.22~~~&0.23~~~&~~~-~~~\\
		& $m_i^{*calc}$&($m_0$)&~~~-&0.24~~~&0.30~~~&1.19~~~\\\hline\hline
	\end{tabular}
	\caption{Parameters obtained from analysis of quantum oscillations measured at $T=2$\,K, $B\parallel c$ and from electronic band structure calculations. 
		\label{FFT_TABLE}}
\end{table}

\begin{figure}[h]
	\includegraphics[width=8cm]{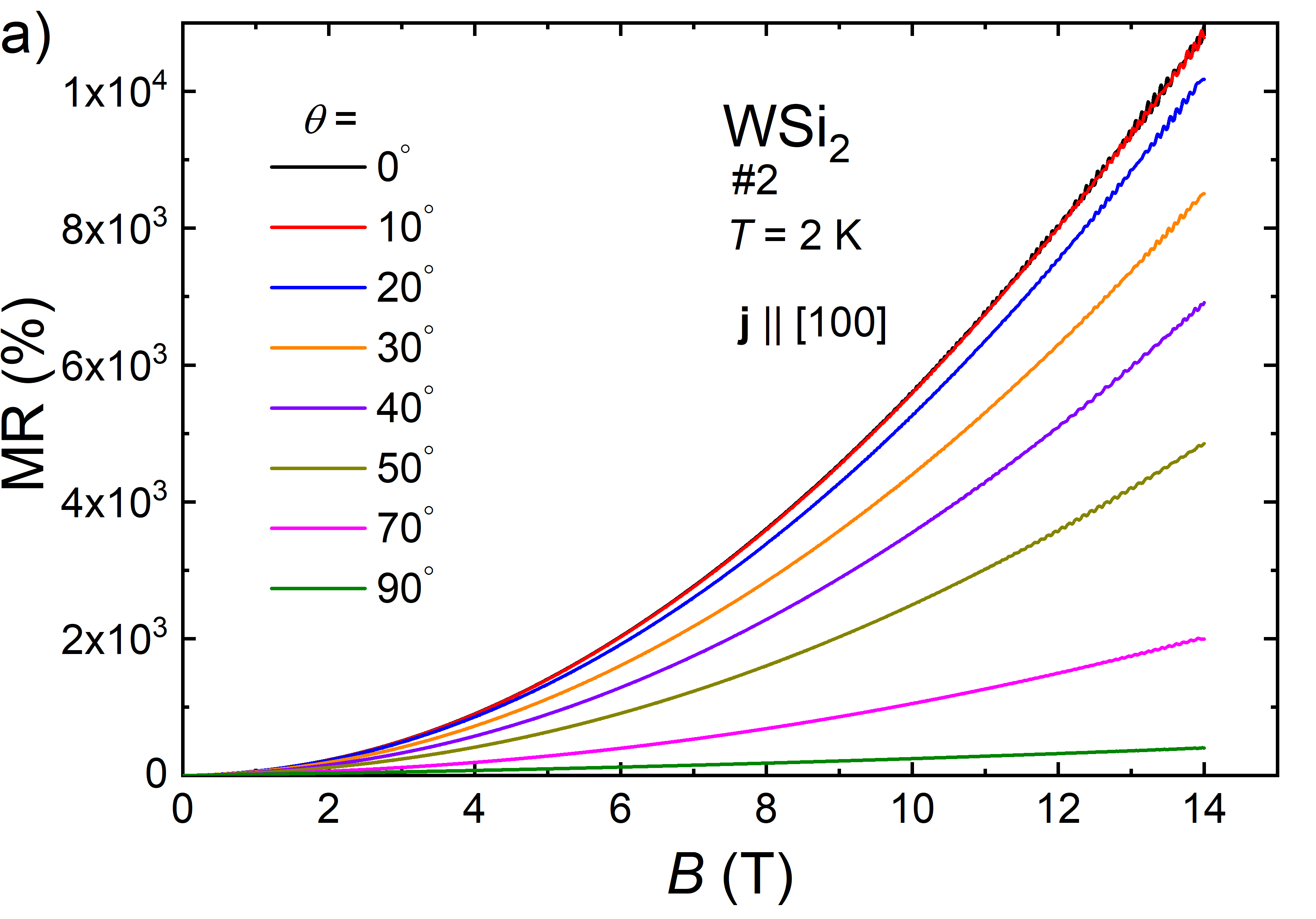}
	\includegraphics[width=8cm]{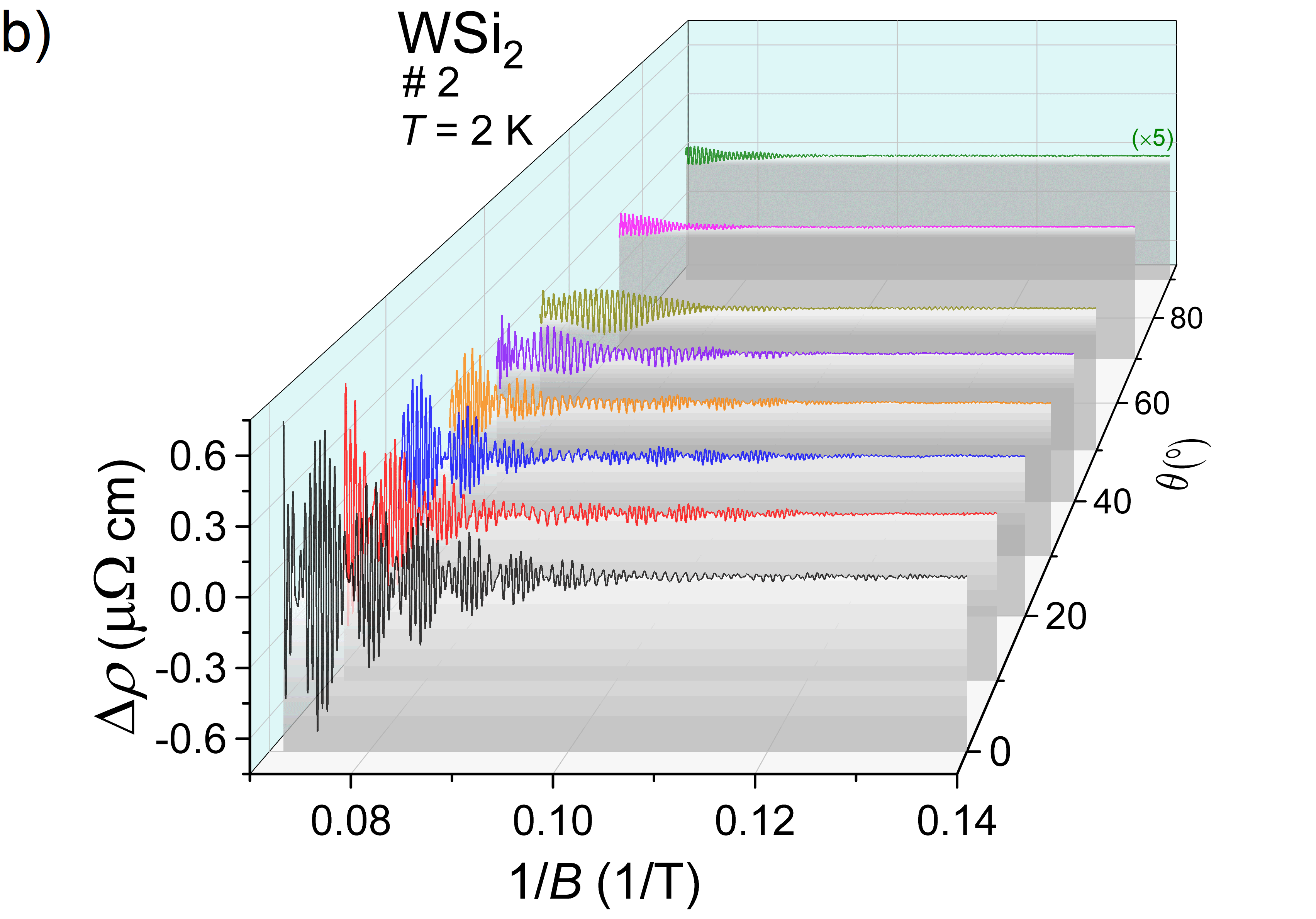}
	\includegraphics[width=8cm]{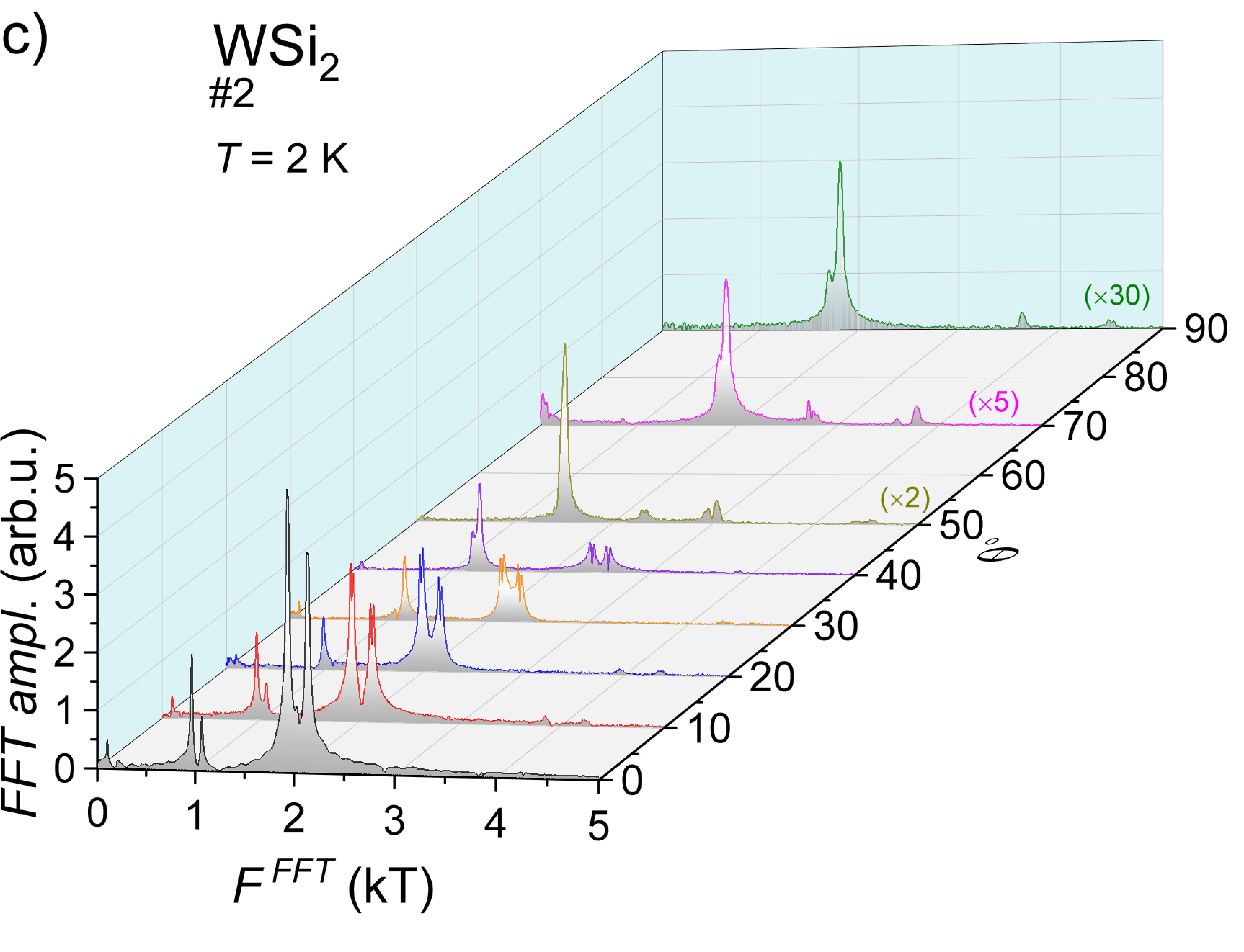}
	\includegraphics[width=8cm]{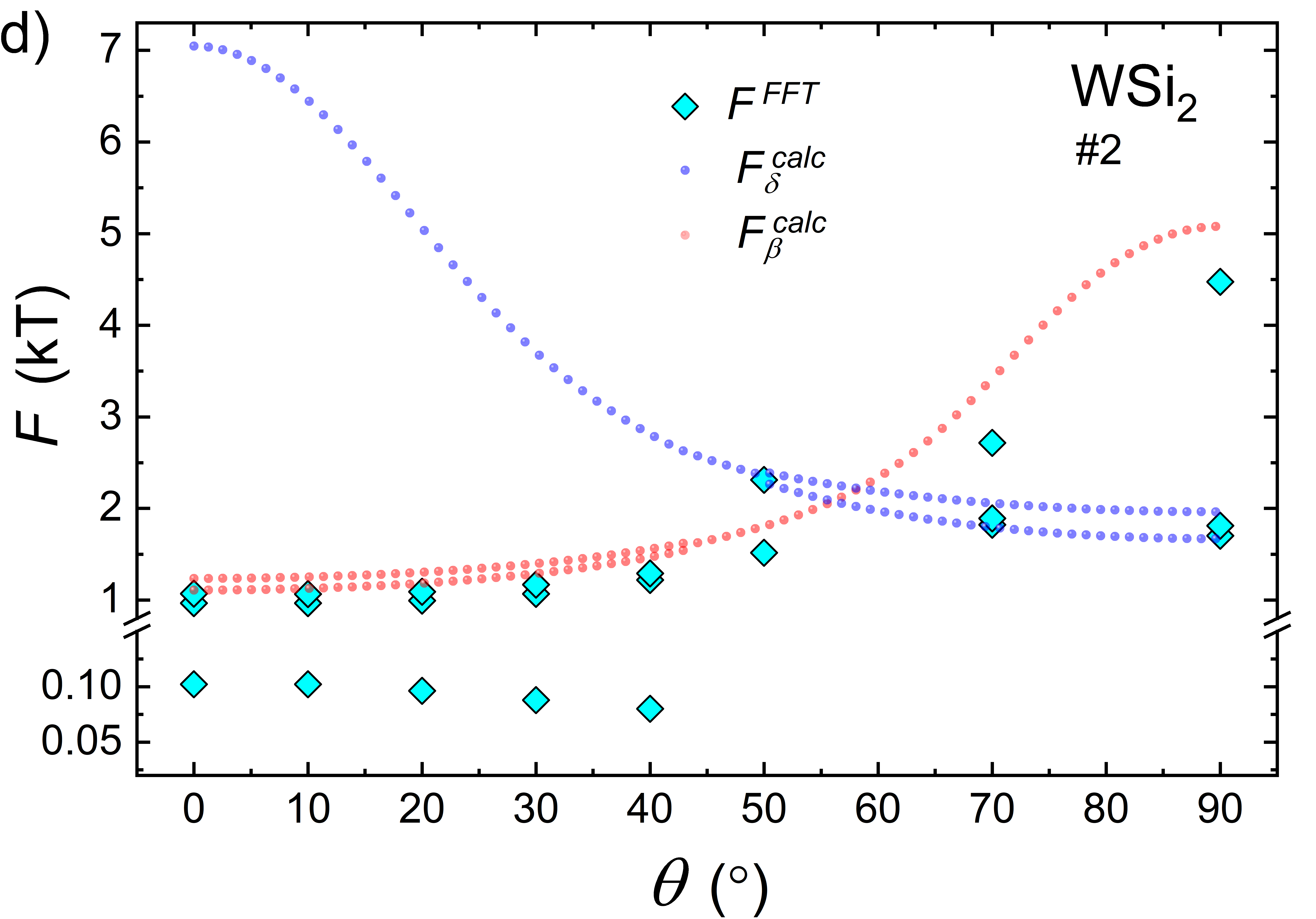}
	\caption{(a) Magnetoresistance of WSi$_2$ (sample $\#2$) measured at $T=2$\,K as a function of magnetic field applied at different angles with respect to the current direction [100]. 
		(b) Oscillating part of the electrical resistivity as a function of inverted magnetic field obtained for the data presented in (a). Data for different angles are shifted for clarity. 
		(c) Fast Fourier transform spectra obtained for the data depicted in (b). 
		(d) Angular dependence of oscillations frequencies obtained from the FFT analysis (diamonds) and frequencies obtained from the first-principle calculations (red circles for the dumbbell-shaped hole-like pocket and blue circles for rosette-shaped electron-like pocket). Frequencies corresponding to the third and fourth harmonics are not shown.
		\label{SdH_deg_WSi2}}
\end{figure}

Combination frequencies are the differences or the sums of two fundamental frequencies or their harmonics. 
One of the evidences that $F_{\beta_2-\beta_1}$ is really a combination of frequencies for both studied materials is that the results of electronic structure calculations show no bands, whose topology can correspond to those experimentally observed frequencies.
Even if one assumes that the Fermi level in WSi$_2$ is slightly shifted, which could lead to the appearance of additional electron-like Fermi pocket at the $P$ point of the BZ, the fact that $F_{\beta_1–\beta_2}$ is the combination frequency instead of frequency originating from hypothetical pocket can be justified by several reasons: (i) according to our theoretical calculations (see Fig.~\ref{SdH_deg_WSi2}d), the bifurcation in $F^{calc}(\theta)$ for the hole-like band disappears at $\theta\!>\!43^{\circ}$, which is in an excellent agreement with the absence of any feature of $F_{\beta_1–\beta_2}$ in the FFT spectra for $\theta\!>\!40^{\circ}$; (ii) shift of Fermi level leads to a poorer agreement between theoretical and experimental results for the $\delta$ pocket; (iii) in the FFT spectra of MoSi$_2$, we also observed small frequency of 99\,T and its origin cannot be attributed to the appearance of additional Fermi pocket at the $P$ point of the BZ, because the bottom of the conduction band at the $P$ point is located at $\sim300$\,meV above the Fermi level (see Fig.~\ref{band_structure}h).  
Combination frequencies can appear, for example, due to the magnetic breakdown effect or magnetic interaction effect\cite{Shoenberg1984} also known in literature as the Shoenberg effect.\cite{Shoenberg1968a,Shoenberg1968}
For WSi$_2$ and MoSi$_2$ the former effect can be excluded because the Fermi surface structure (see Fig.~\ref{band_structure}c,d) shows no orbits being very close to each other, thus resistant to magnetic breakdown, especially when the applied magnetic field is parallel to the [001] crystallographic direction.

On the other hand, the effect of magnetic interactions have been intensively studied in single crystals of many highly pure elements (beryllium, silver, gold) more than half a century ago.\cite{Shoenberg1984} 
Usually, this effect can be observed only at relatively low temperatures below $\sim5$\,K.
The combination frequencies due to magnetic interactions can appear not only from the frequencies belonging to two different Fermi pockets, but  also can steam from two frequencies associated to opposite extrema of the same Fermi surface pocket, as it takes place in case of studied materials.   
It is quite unusual that we observed the combination frequency at such high temperatures ($T=12$\,K for WSi$_2$ and $T=25$\,K for MoSi$_2$, the highest temperature at which the magnetic interactions effect has been observed in any material). 
Moreover, $F_{\beta_2-\beta_1}$ is the only frequency observed in the FFT spectrum of MoSi$_2$ at $T=25$\,K. 
A similar effect has been noticed for the single crystal of silver, but at much lower temperatures,\cite{Joseph1965} and its origin has remained still unexplained.
The magnetic interactions effect leads to the special shape of the oscillations, which differs from those predicted by the Lifshitz-Kosevich (L.-K.) theory,\cite{Shoenberg1984} according to which the oscillating component of electrical resistivity can be described by:
\begin{equation}
	\begin{matrix*}[l]
		\Delta\rho \simeq \frac{5}{2} \sum\limits_{i} A_i\sqrt{\frac{B}{2p_iF_i}}R_{T,i}R_{\rm{D},i}R_{S,i}\cos\bigg(2\pi p_i\bigg(\frac{F_i}{B}+\gamma_i\bigg)\bigg),\\
		R_{T,i}=(p_i\lambda m^*_iT/B)/\sinh(\lambda m^*_iT/B),\\
		R_{\rm{D},i}=\exp(-p_i\lambda m^*_iT_{\rm{D,i}}/B),\\
		R_{S,i}=\cos(p_i\pi g_im^*_i/(2m_0)),\\
		\gamma_i=1/2-\phi_{B,i}/2\pi\pm \delta
	\end{matrix*}
	\label{full_LK_eq}
\end{equation}
where $A_i$ is a scaling coefficient, $R_i(T)$ corresponds to temperature damping factor, $m^*_i$ is the effective mass of carriers, $\lambda$ is a constant which equals to $2\pi^2k_Bm_0/e\hbar\:(\approx14.7\;$T/K), $p_i$ is the number of harmonic; $R_{\rm{D},i}$ is the so-called Dingle factor, which is related to the electron scattering, $T_{\rm{D,i}}$ is the Dingle temperature; $R_{S,i}$ denotes the spin reduction factor, $g_i$ is the Land\'e $g$-factor, $\gamma_i$ stands for the phase shift of quantum oscillations, $\phi_{B,i}$ is Berry phase and $\delta=\pm 1/8$ (sign before $1/8$ depends on the type of carriers (electrons or holes) and on the kind of extremum orbit (minimal or maximal). 
Therefore, the saw-tooth shape of quantum oscillations at $T=25$\,K (see lower inset to Fig.~\ref{SdH_analysis}a) is one more indication that the magnetic interactions effect, and not a separate Fermi sheet, is the source of the smallest frequency in the FFT spectrum. 

\begin{figure}[h] 
	\includegraphics[width=8cm]{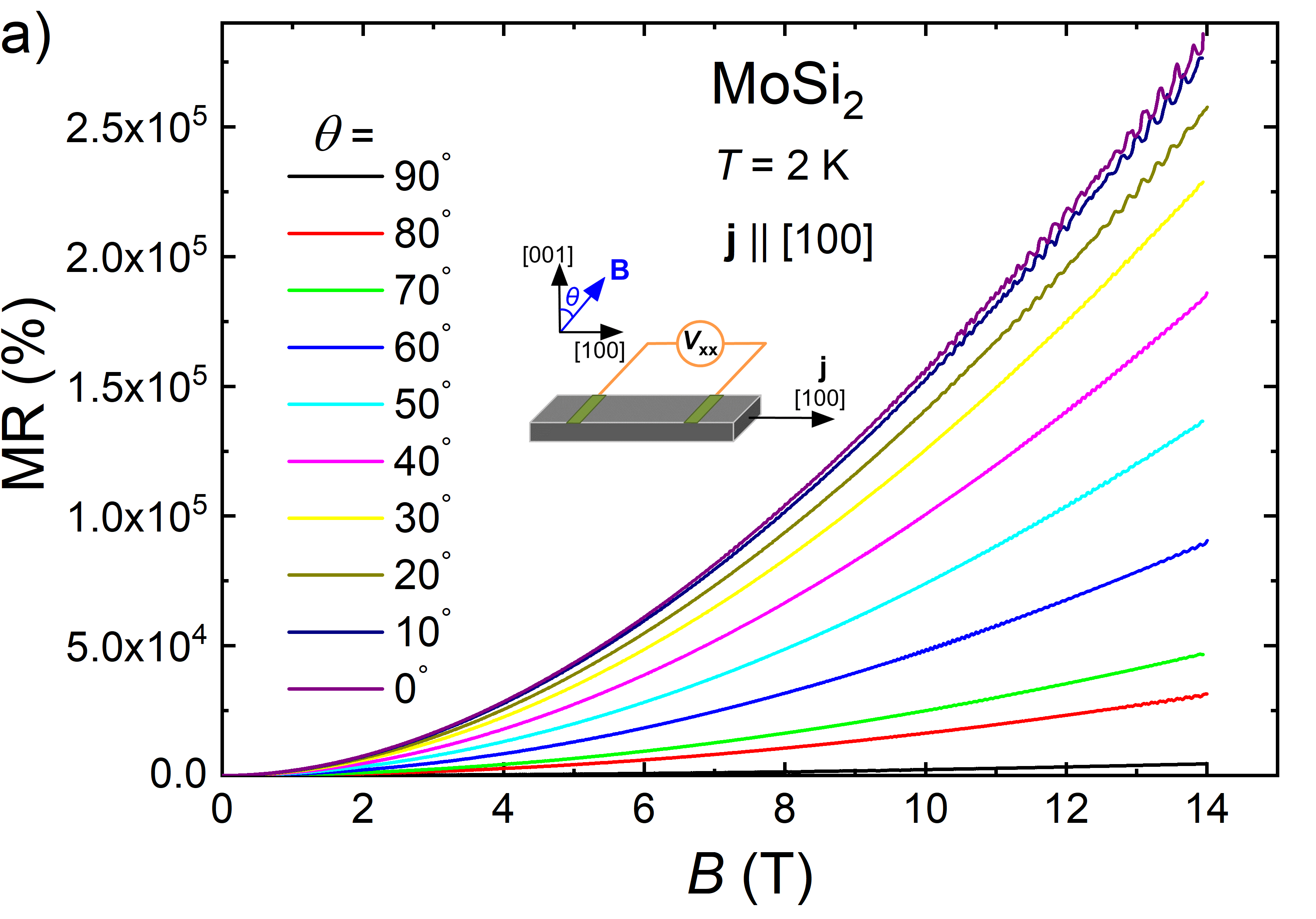}
	\includegraphics[width=8cm]{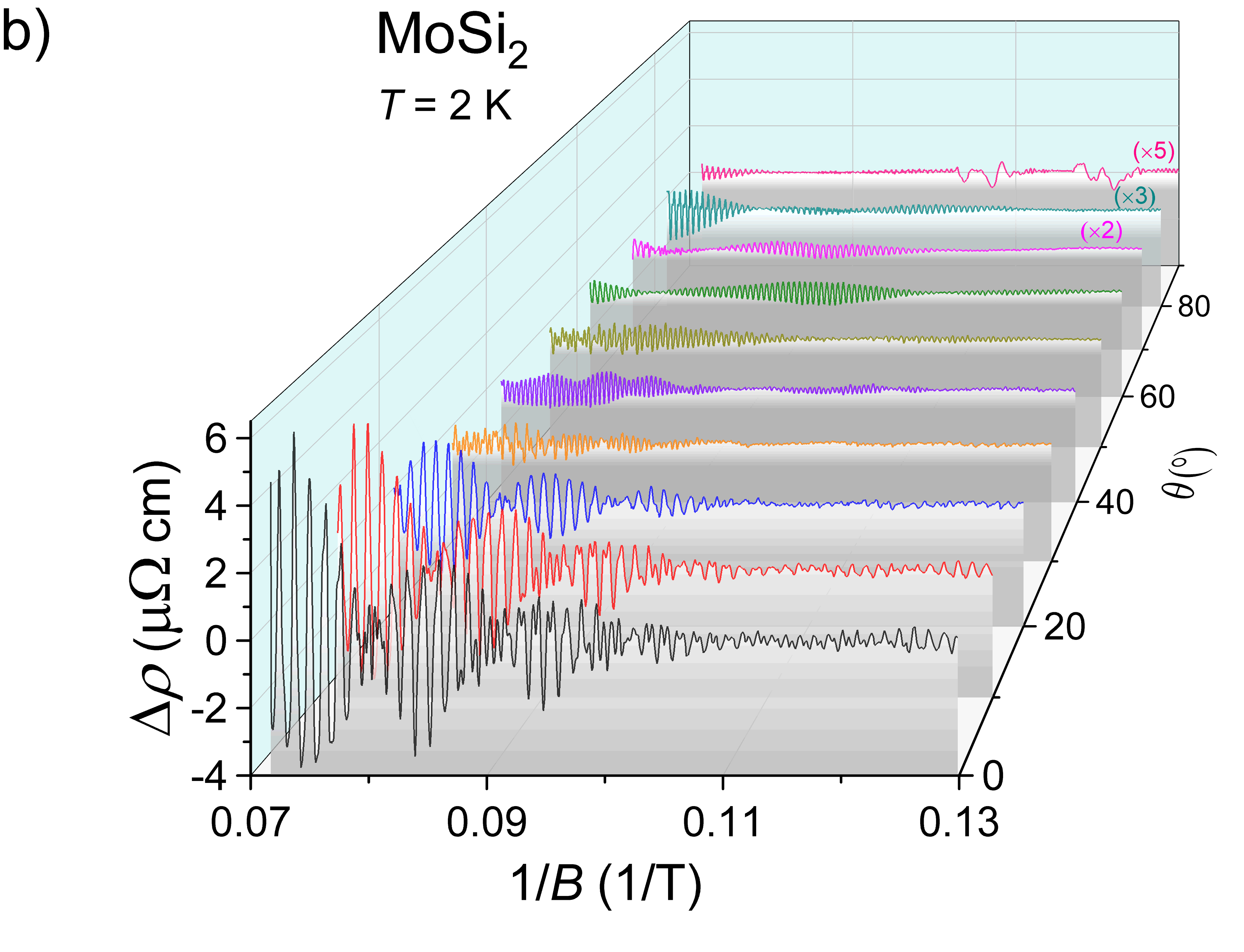}
	\includegraphics[width=8cm]{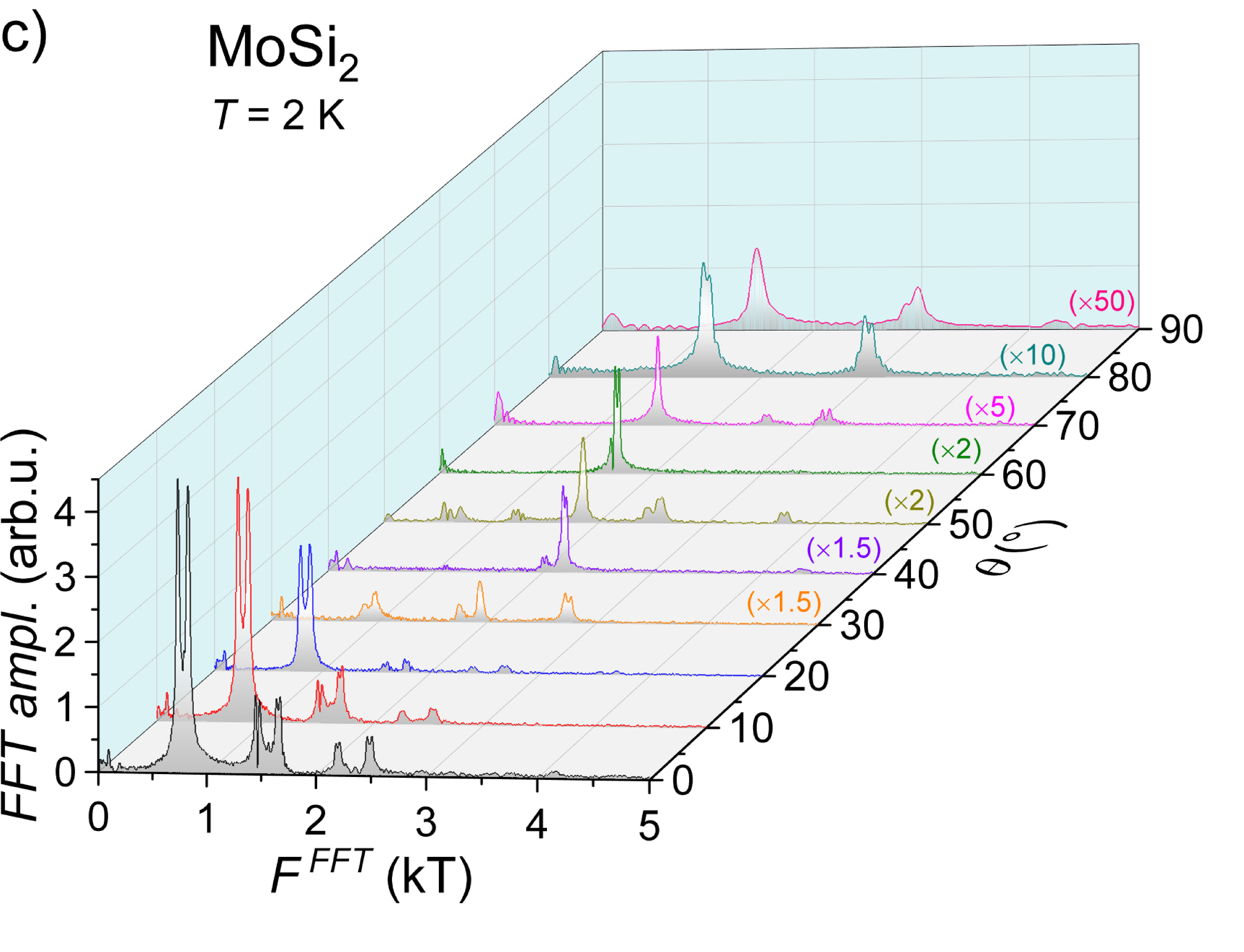}
	\includegraphics[width=8cm]{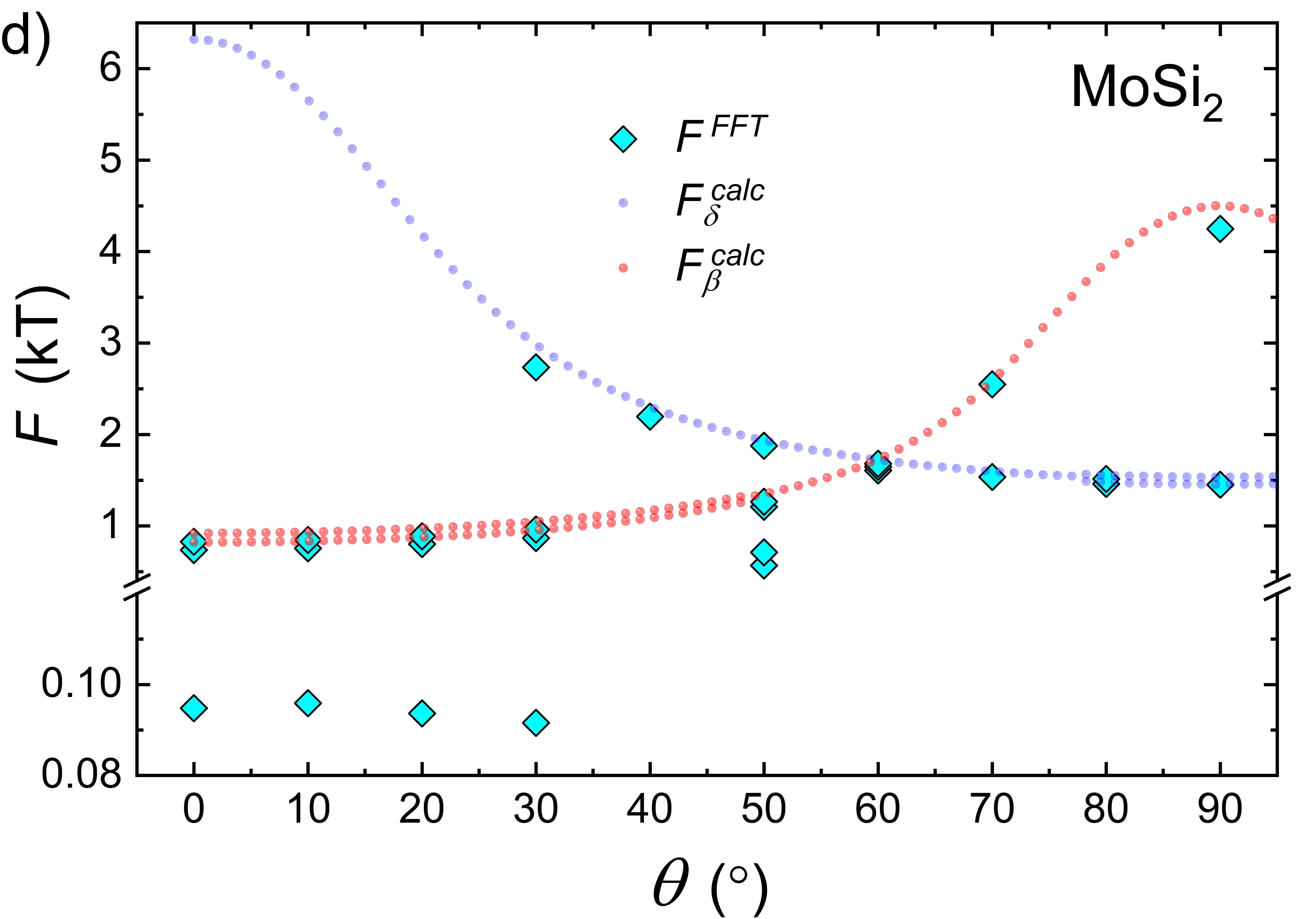}
	\caption{(a) Magnetoresistance of MoSi$_2$ measured at $T=2$\,K as a function of magnetic field applied at different angles with respect to the current direction. 
		Inset schematically shows the measurement geometry. 
		(b) Oscillating part of the electrical resistivity as a function of inverted magnetic field obtained for the data presented in (a). 
		(c) Fast Fourier transform spectra obtained for the data depicted in (b).
		(d) Angular dependence of oscillations frequencies obtained from the FFT analysis (diamonds) and frequencies obtained from the first-principle calculations (red circles for the dumbbell-shaped hole-like pocket and blue circles for rosette-shaped electron-like pocket). Frequencies corresponding to the second and third harmonics are not shown.
		\label{SdH_deg_MoSi2}}
\end{figure}

In comparison to the literature data, where the dHvA oscillations are analyzed, we obtained fairly good quantitative and qualitative agreement, apart from two things: (i) we observed the combination frequencies $F_{\beta2-\beta1}\sim100$\,T, which have not been previously reported for either of two materials studied\cite{Ruitenbeek1987,Matin2018,Mondal2020}, and (ii) in contrast to the literature reports,\cite{Mondal2020,Ruitenbeek1987} $F_\delta$ (when $B\parallel c$) was missing in the FFT spectra of both compounds (see Fig.\ref{SdH_analysis}b,d). 
The $F_{\delta}$ frequency was not observed for $B\parallel[001]$ due to its larger effective masses compared to the effective masses for $\beta$ pocket. 
Based on this, we suppose this frequency could be observed at temperatures smaller than 2\,K, the lowest temperature at which our experiments have been performed.  
Nevertheless, due to the anisotropy of the effective mass of the $\delta$ Fermi pocket, the oscillations frequency related to this pocket can be easyly distinguished from the frequency spectrum at $\theta\geq50^{\circ}$ for WSi$_2$ (see Fig.~\ref{SdH_deg_WSi2}) and at $\theta\geq30^{\circ}$ for MoSi$_2$ (see Fig.~\ref{SdH_deg_MoSi2}) even at $T=2$\,K. 

For WSi$_2$, we ascribed $\beta_1$ and $\beta_2$ frequencies to the extrema cross-sections of the dumbbell-like pocket (see Fig.~\ref{band_structure}c). 
In this case, the interpretation fully agrees with that previously reported in Ref.\onlinecite{Mondal2020}. 
In turn, for MoSi$_2$, our frequency interpretation differs from that reported in Ref.~\onlinecite{Matin2018}, but it is in a full agreement with that showed in Ref.~\onlinecite{Ruitenbeek1987}. 
In the first work, the authors ascribed two fundamental frequencies (which we denoted as $\beta_1$ and $\beta_2$) to two separate Fermi pockets; however, here, we confirmed by means of both the electronic structure calculations and the angle dependent quantum oscillations analysis (see Fig.~\ref{SdH_deg_MoSi2}d), that these frequencies originate from a single Fermi pocket.

Interestingly, at $T<8$\,K the amplitudes of the second harmonic frequencies of WSi$_2$ are larger than the amplitudes of the corresponding fundamental frequencies (see Fig.~\ref{SdH_analysis}d). 
This behavior can be related either to the magnetic interactions effect or to the Zeeman spin-splitting effect.\cite{Shoenberg1984} 
However, we can assume that the latter effect is more likely due to the following reasoning:  
at $T\geq8$\,K, we observe the reversed ratio of amplitudes (if compared to data at $T<8$\,K), i.e., the amplitudes of fundamental frequencies are larger than the harmonic ones. 
In the same temperature range, we noticed that the combination frequencies of the oscillations are still distinguished in the FFT spectra (see Fig.~\ref{SdH_analysis}b,d) and the peak splitting is not more noticeable (see upper insets to Fig.~\ref{SdH_analysis}a,c, and description below). 
Based on this, the magnetic interactions effect can be excluded as the dominant source of the observed amplitude ratio.
We additionally proved that quantum oscillations in WSi$_2$ reveal Zeeman spin-splitting effect by performing their FFT analysis for different intervals of magnetic fields (for details see Supplementary Information). It was found that the ratio of the fundamental oscillation amplitude to the amplitude of the second harmonic oscillation $(A_{\beta_i}/A_{2\beta_i})$ become larger when the magnetic field interval is narrowed: for low fields $A_{\beta_i}/A_{2\beta_i}\!>\!1$ and for strong magnetic fields $A_{\beta_i}/A_{2\beta_i}\!<\!1$. 
Earlier, even more pronounced influence of the Zeeman effect on the second harmonic oscillations was reported for the Dirac system Pb$_{0.83}$Sn$_{0.17}$Se.\cite{Orbanic2017} 
For that particular material the harmonics oscillations completely disappear with the change of magnetic field interval, for which FFT was performed.

The effective mass of carriers is related to the curvature of electronic bands, thus it can acquire different values for a single Fermi pocket. 
Due to the fact that electron-like Fermi pockets have two extremal cross-sections, which are fully detectable in the observed quantum oscillations, it is possible to calculate effective masses for both of them, ($m^*_{\beta_1}$ and $m^*_{\beta_2}$). 
As it is shown in the inset to Fig.~\ref{SdH_analysis}b and Fig.~\ref{SdH_analysis}d, all $m^*_i$ were obtained from the fitting of temperature dependences of FFT amplitudes to the temperature damping factor, $R_i(T)$ of the Lifshitz-Kosevich formula (Eq.\ref{full_LK_eq}).\cite{Shoenberg1984}
As the FFT amplitudes and not oscillations amplitudes were used during the effective mass estimation, we changed $B$ into $B_{e\!f\!f}$ in the formula for $R_i(T)$ (Eq.\ref{full_LK_eq}). 
$B_{e\!f\!f}$ is the reciprocal of average inverse field from the window where the FFT was performed, in our case $B_{e\!f\!f}=2(1/14+1/8)=10.18$\,T. 
The obtained effective masses are listed in Table~\ref{FFT_TABLE}.  
For MoSi$_2$ and WSi$_2$ the effective masses are similar, confirming the close resemblance between the electronic structures of both compounds. 
Importantly, the experimentally determined effective masses are almost identical to those calculated from first principles theory, $m_i^{*calc}$ (see Table~\ref{FFT_TABLE}).
The reliable determination of effective mass corresponding to $F_{\beta_2-\beta_1}$ frequencies was not possible, because the temperature dependences of their FFT amplitude are non-monotonic.
This behavior can be either related to the temperature dependence of the Fermi surface\cite{Wu2015} or to the Shoenberg effect.\cite{Shoenberg1984,Shoenberg1962} 
As the observed non-monotonic properties relate to the combination frequencies, one can assume that the Shoenberg effect is responsible for that behavior.  

The splitting of peak and valley, shown by blue and red arrows in the inset to Fig.~\ref{SdH_analysis}a for MoSi$_2$ and in the inset to Fig.~\ref{SdH_analysis}c for WSi$_2$, can be attributed to the spin-splitting effect. 
This effect has been reported for several topologically trivial\cite{Shoenberg1984} and non-trivial\cite{Liu2016,Matusiak2017} materials. 
For the latter group it can lead to unusual phenomena such as the anomalous Hall effect.\cite{Sun2020b}
The splitting is the most pronounced at $T=2$\,K. 
At higher temperatures it gradually smears due to the thermal broadening of Landau levels, disappearing completely around $T=8$\,K for both materials.
The Land{\'{e}} $g$-factor, $g_i$, which is attributed to the measure of the strength of the Zeeman effect, can be calculated using the harmonic ratio method.\cite{Shoenberg1984,Gold1976}
The FFT spectrum of MoSi$_2$ contains well pronounced peaks corresponding up to the 3rd harmonics of $F^{FFT}_{\beta_1}$ and $F^{FFT}_{\beta_2}$. 
It allows us to estimate the $g$-factors, knowing only the effective masses and the amplitudes of the particular frequencies.
It should be noted that ambiguity in the $g$-factor determination from the quantum oscillations exists. 
In most cases analysis of quantum oscillations gives the lowest limit of the $g$-factor, ($g_{0,i}$),\cite{Shoenberg1984} which is often omitted in the research articles. 
According to theory,\cite{Shoenberg1984} the $g$-factor may be equal to any value calculated from the following equation:
\begin{equation}
	g_i=\frac{2r}{m^*_i/m_0}\pm g_{i,0},	
	\label{g-factor_eq}	
\end{equation}
where $r$ is an integer number. 
We obtained the following values of $g_{0,i}$ for the $\beta$ Fermi sheet of MoSi$_2$: $g_{0,\beta_1}=1.8$ and $g_{0,\beta_2}=2.7$. 
These values differ slightly from the $g$-factor of free electron of 2. 
Using Eq.~\ref{g-factor_eq}, we found that for MoSi$_2$, $g_{\beta_1}$ could be equal to one value from the series: 1.8, 9.8, 6.2, 17.8, 14.2 etc., and $g_{\beta_2}$ could be equal to one value from the series: 2.7, 10.4, 5, 18.1, 12.7 etc. 
It is possible to reduce the number of putative values of $g$-factors, however, more information about the phase shift of quantum oscillations or experiments in higher magnetic fields are required.\cite{Shoenberg1984} 
For WSi$_2$ the FFT spectra are not so rich in detail and only the second harmonics were noticed, thus we used the second harmonic variant of the harmonic ratio method\cite{Gold1976} to calculate $g_i$. 
This variant of the method requires the knowledge of the Dingle temperature value. 

\begin{figure}[h]
	\includegraphics[width=8cm]{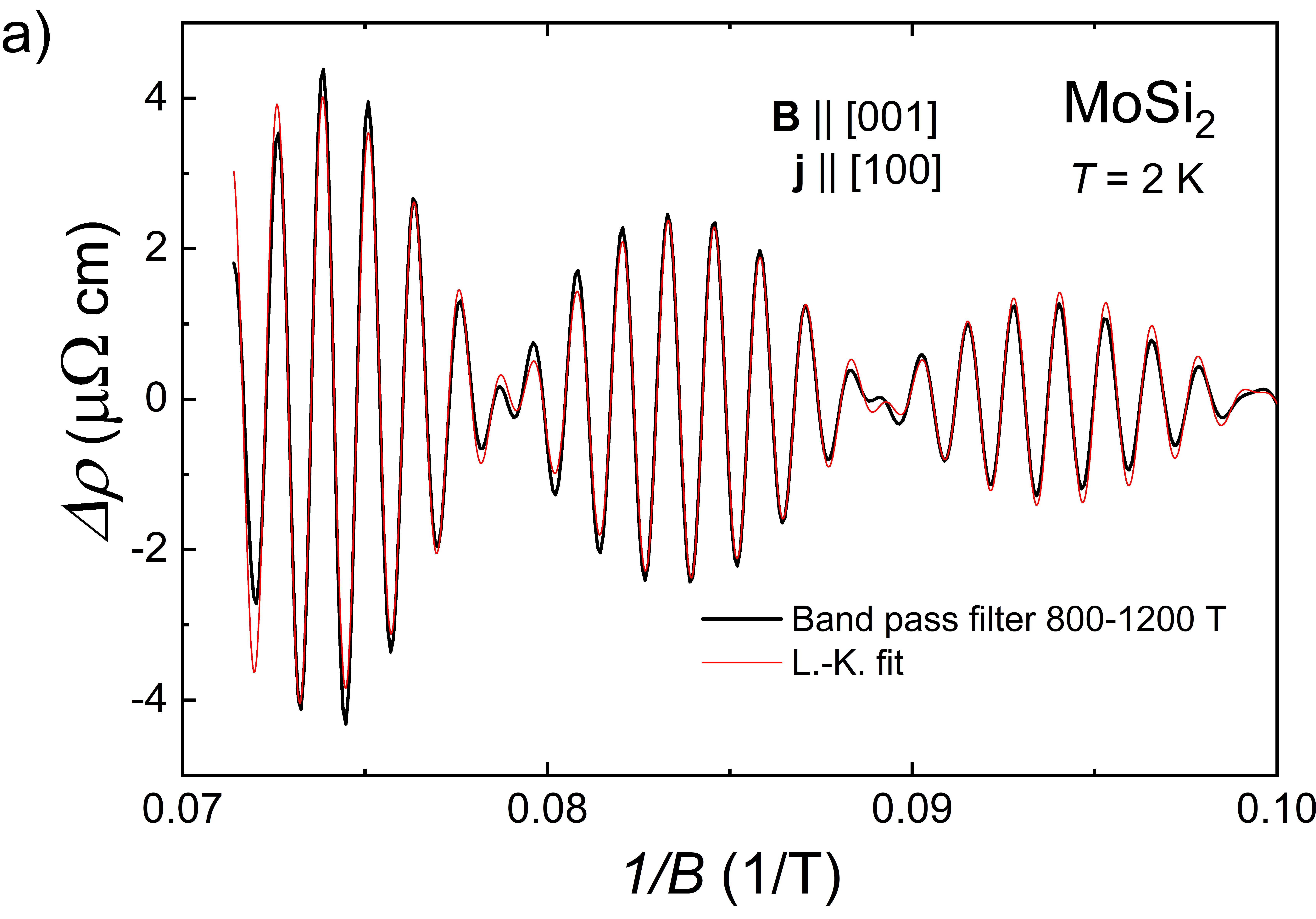}
	\includegraphics[width=8cm]{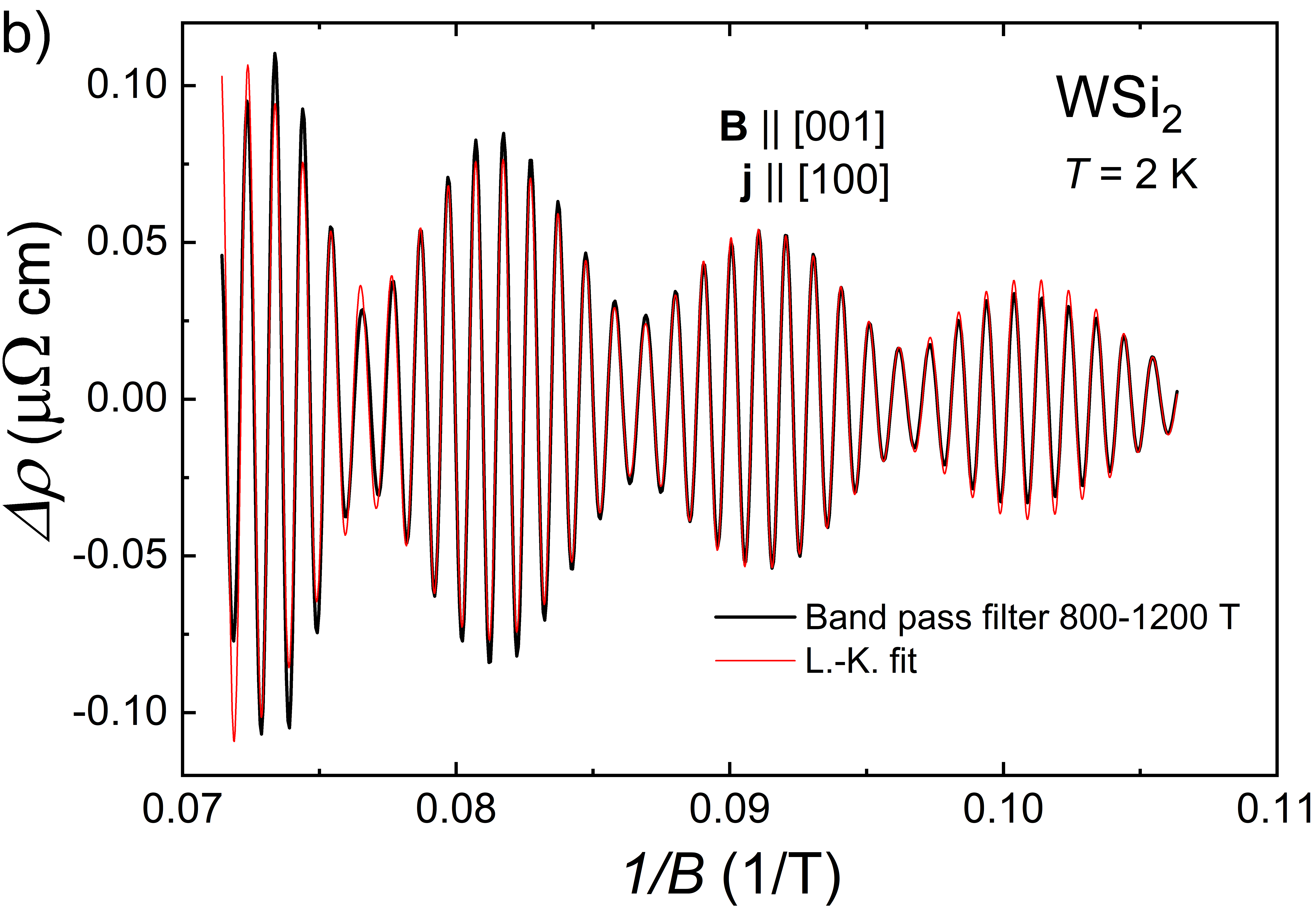}
	\caption{Oscillating part of the electrical resistivity of MoSi$_2$ (a) and sample $\#2$ of WSi$_2$ (b) as function of inverted magnetic field obtained after band-pass FFT filtering performed to isolate the signal originated from the $\beta$ Fermi pocket. Red solid lines show the fits to the Lifshitz-Kosevich formula (Eq.~\ref{full_LK_eq}). 
		\label{SdH_filtering}}
\end{figure}

The Dingle temperatures were obtained from the direct fitting of the Lifshitz-Kosevich formula (Eq.~\ref{full_LK_eq}) to the experimental data.
In order to decrease the amount of fitting parameters, the band-pass filter was used and the oscillations related to the $\beta_1$ and $\beta_2$ frequencies were isolated for both compounds (see Fig.~\ref{SdH_filtering}). 
To avoid the mutual dependence between fitting parameters, we fixed the values of effective masses to the priorly obtained values (see Table~\ref{FFT_TABLE}) and introduced a new fit parameter $C_i=A_iR_{s,i}$.
The least-square fitting yielded the following Dingle temperatures: $T_{\rm{D,\beta_1}}=10.2$\,K and $T_{\rm{D,\beta_2}}=13.1$\,K for MoSi$_2$; $T_{\rm{D,\beta_1}}=9.2$\,K and $T_{\rm{D,\beta_2}}=8$\,K for WSi$_2$.
For MoSi$_2$, the obtained values of $T_{\rm D}$ are larger than those reported in Ref.~\onlinecite{Matin2018}, and for WSi$_2$ $T_{\rm D}$ values agree well with those reported in Ref.~\onlinecite{Mondal2020}. 
Using the obtained $T_{\rm{D}}$, the lowest limit of the $g$-factors for WSi$_2$ were estimated to be $g_{0,\beta_1}=4.5$ and $g_{0,\beta_2}=4.3$. 
These values are more than twice as large as those obtained for MoSi$_2$, originating from the stronger SOC effect in WSi$_2$.
The knowledge of $T_{\rm{D}}$ allows us also to calculate quantum scattering lifetime, $\tau_{Q,i}$. 
We used the following formula $\tau_{Q,i}=\hbar/(2\pi\!k_BT_{\rm{D},i}$), which gave the following values: $\tau_{Q,\beta_1}=1.2\times10^{-13}$\,s and $\tau_{Q,\beta_2}=9.3\times10^{-14}$\,s for MoSi$_2$; $\tau_{Q,\beta_1}=1.3\times10^{-13}$\,s and $\tau_{Q,\beta_2}=1.5\times10^{-13}$\,s for WSi$_2$. 
Taking into account effective mass and scattering time, the quantum mobility can be obtained using the relation $\mu_{Q,i}=e\tau_{Q,i}/m^*_i$, we got $\mu_{Q,\beta_1}=839\,\rm{cm^{2}V^{-1}s^{-1}}$ and $\mu_{Q,\beta_2}=628\,\rm{cm^{2}V^{-1}s^{-1}}$ for MoSi$_2$; $\mu_{Q,\beta_1}=1057\,\rm{cm^{2}V^{-1}s^{-1}}$ and $\mu_{Q,\beta_2}=1163\,\rm{cm^{2}V^{-1}s^{-1}}$ for WSi$_2$.

Another kind of mobility $\mu_{H}$, the so-called classical mobility, can be determined from the analysis of Hall effect data.   
The principal difference between classical mobility and quantum mobility is that the former relates to the large angle scattering, whereas the latter relates to both small and large angle scattering.\cite{Narayanan2014} 
The relation between these mobilities, $r_i=\mu_{H,i}/\mu_{Q,i}$, is frequently used as a criterion of the strength of the backscattering suppression.\cite{Kumar2017c} 
From the analysis of Hall effect data of MoSi$_2$ (see Supplementary Information), we found that $\mu_{H,\beta}=6.42\!\times\!10^4\,\rm{cm^{2}V^{-1}s^{-1}}$, and $r_{\beta}=102$ (the value of $\mu_{Q,\beta_2}$ was used to calculate $r_{\beta}$). 
This value of $r_{\beta}$ is of the same order of magnitude as that reported for MoSi$_2$ in Ref.~\onlinecite{Matin2018}, however it is two orders of magnitude smaller than those values reported for archetypal Dirac semimetal Cd$_3$As$_2$ in Ref.~\onlinecite{Liang2014} or for WP$_2$ in Ref.~\onlinecite{Kumar2017c}.

\begin{table}
	\centering
	\begin{tabular}{*{7}{c}} \hline\hline 
		{Compound}&&Fermi sheet cross-section~~~~&$C_i$~~~~&$\gamma_i$~~~~&$\delta_i$~~~~&$\phi_{B,i}$\\\hline
		MoSi$_2$&~~~~~~~~~~&$\beta_1$&128.77~~~~&0.14~~~~&-1/8~~~&$0.47\pi$\\
		&~~~~~~~~~~&$\beta_1$&-128.77~~~~&-0.36~~~~&-1/8~~~&$1.47\pi$\\
		&~~~~~~~~~~&$\beta_2$&409.24~~~~&-0.11~~~~&1/8~~~&$1.47\pi$\\
		&~~~~~~~~~~&$\beta_2$&-409.24~~~~&0.39~~~~&1/8~~~&$0.47\pi$\\\hline
		WSi$_2$&~~~~~~~~~~&$\beta_1$&3.12~~~~&0.06~~~~&-1/8~~~&$0.45\pi$\\
		&~~~~~~~~~~&$\beta_1$&-3.12~~~~&-0.44~~~~&-1/8~~~&$1.63\pi$\\
		&~~~~~~~~~~&$\beta_2$&1.21~~~~&-0.3~~~~&1/8~~~&$1.85\pi$\\
		&~~~~~~~~~~&$\beta_2$&-1.21~~~~&0.2~~~~&1/8~~~&$0.85\pi$\\\hline\hline
	\end{tabular}
	\caption{Phase shift of quantum oscillations obtained from the L.-K. analysis of the SdH oscillations in MoSi$_2$ and WSi$_2$. $C_i=A_iR_{s,i}$; $\gamma_i$ is a phase shift; $\delta_i$ stands for the phase shift correction; $\phi_{B,i}$ is the Berry phase.
		\label{phase_shift}}
\end{table}

An additional parameter which can be also extracted from the L.-K. fit is the phase shift, but its unambiguous determination is not possible without knowledge of the sign of $R_{s,i}$ damping factor.\cite{Shoenberg1984} 
$R_{s,i}$ could be in the range from -1 to 1, which leads to two possible values of Berry phase that differ from each other by a factor of $\pi$ (see Table~\ref{phase_shift}). 
This makes it impossible to distinguish, without the knowledge of the value of spin-splitting factor, if Berry phase is trivial or non-trivial.
For both compounds for the case of negative $R_{S,i}$ the total phase shifts $\gamma_{\beta_1}$ and $\gamma_{\beta_2}$ are much closer to the values reported in literature\cite{Matin2018,Mondal2020} than the values $\gamma_{\beta_1}$ and $\gamma_{\beta_2}$ obtained for the case of positive $R_{S,i}$.        


According to the Onsager relation $\big(F_i=\frac{hS_i}{4\pi^2e}\big)$ frequency of quantum oscillations is proportional to the area $S_i$ of the Fermi pocket cross-section.\cite{Shoenberg1984} 
This simple relation plays an essential role in the investigation of  Fermi surface topology. 
The synergy between theoretical calculations of Fermi surface and experimentally observed quantum oscillations allows for characterization of the Fermi surface with high accuracy. 
Theoretically calculated frequencies as functions of angle $\theta$ are shown as full circles in Fig.~\ref{SdH_deg_MoSi2}d and Fig.~\ref{SdH_deg_WSi2}d for particular Fermi sheets of MoSi$_2$ and WSi$_2$, respectively.
To make the complete experimental identification of the Fermi surface of MoSi$_2$ and WSi$_2$, we studied the angle-dependent magnetoresistance (Fig.~\ref{SdH_deg_MoSi2}a and Fig.~\ref{SdH_deg_WSi2}a), each sample was rotated in a magnetic field in a way sketched in the inset to Fig.~\ref{SdH_deg_MoSi2}a. 
The quantum oscillations were observed in the entire range of $\theta$, however their amplitudes decrease with increasing $\theta$ (see Figure~\ref{SdH_deg_MoSi2}b and Fig.~\ref{SdH_deg_WSi2}b). 
The obtained FFT spectra are shown in Fig.~\ref{SdH_deg_MoSi2}c and Fig.~\ref{SdH_deg_WSi2}c, and the angular dependence of the extracted frequencies are presented as empty symbols in Fig.~\ref{SdH_deg_MoSi2}d and Fig.~\ref{SdH_deg_WSi2}d where they are compared to those theoretically calculated. 
For both compounds, $F^{FFT}_{\beta}(\theta)$ and $F^{FFT}_\delta(\theta)$ qualitatively represent the theoretical behavior of dumbbell-shaped and rosette-shaped Fermi pockets, respectively. 

We observed frequencies originating from the $\beta$ pocket in the entire covered range of the angle $\theta$. 
This made it possible to calculate the volume of that pocket ($V_{F,\beta}$) with good accuracy. 
This volume is proportional to the carrier concentration, $n_{F, i}=V_{F,i}/(4\pi^3)$. 
For the sake of simplicity, we assume that $\beta$ pocket is cylinder-shaped with the base area $S_{F,\beta_2}=0.10194\,$\r{A}$^{-2}$ and $S_{F,\beta_2}=0.08047\,$\r{A}$^{-2}$ for WSi$_2$ and MoSi$_2$, respectively.
Then the area of the cross-section which is perpendicular to the base of the cylinder equals $S_{F,\beta}(\theta\!=\!90^{\circ})=0.42714\,$\r{A}$^{-2}$ and $S_{F,\beta}(\theta\!=\!90^{\circ})=0.4054\,$\r{A}$^{-2}$ for WSi$_2$ and MoSi$_2$, respectively. 
From the above areas, the following Fermi wave vectors were determined: $k_{F,\beta_2}=0.18018\,$\r{A}$^{-1}$
and $k_{F,\beta}=0.592657\,$\r{A}$^{-1}$ for WSi$_2$;
$k_{F,\beta_2}=0.16008\,$\r{A}$^{-1}$ and
$k_{F,\beta}=0.63309\,$\r{A}$^{-1}$ for MoSi$_2$.
Using the above values, we calculated volumes of pockets and next the corresponding carrier concentrations which are $n_{F,\beta}=9.75\times10^{20}\rm{cm}^{-3}$ and $n_{F,\beta}=8.21\times10^{20}\rm{cm}^{-3}$ for WSi$_2$ and MoSi$_2$, respectively.  
These values are very similar to those obtained from the electronic structure calculations (see Table~\ref{FFT_TABLE}), and for MoSi$_2$ it is very close to that obtained from Hall effect analysis (see Supplementary Information).

In addition to the verification of the Fermi surface, studies of the angle-dependent quantum oscillations of MoSi$_2$ revealed the spin-zero effect. 
At $\theta=40^{\circ}$, FFT spectrum of MoSi$_2$ demonstrates rather unusual features, namely the amplitudes of the fundamental oscillations with frequencies $F_{\beta_1}$ and $F_{\beta_2}$ become negligibly small, whereas their second harmonics are well pronounced.
According to the Lifshitz-Kosevich formula (Eq.~\ref{full_LK_eq}), such scenario can take place only if the spin reduction factor $R_{S,i}$  equals to zero. 
To fulfill this requirement, $p_ig_im^*_i/m_0$ should be an odd integer ($p_i=1$ for fundamental frequency) and in such case the second-harmonic ($p_i=2$) $p_ig_im^*_i/m_0$ will be an even integer, which leads to $R_{S,i}=\pm1$ and amplitude of the second-harmonic frequency will be well pronounced.

The spin-zero effect is not a commonly-observed phenomenon. It has also been noticed in highly pure simple metals like copper, platinum and gold,\cite{Shoenberg1984} and only in a few topological semimetals, such as WTe$_2$\cite{Bi2018} and ZrTe$_5$\cite{Wang2018g}.   
The spin-zero effect can be used in the estimation of the lowest limit of the $g$-factor.
The amplitude of the fundamental frequency disappears if only $R_{S,i}=0$, which means that $g_i=(2r+1)/(m^*/m_0)$.\cite{Shoenberg1984}
As we obtained very good agreement between $m^{*}_i$ and $m^{*calc}_i$ at $\theta=0^{\circ}$ (see Table~\ref{FFT_TABLE}), we used the calculated effective masses at $\theta=40^{\circ}$ ($m^{*calc}_{\beta_1}=0.38\,m_0$ and $m^{*calc}_{\beta_2}=0.39\,m_0$) to estimate $g$-factors.
The obtained $g$-factors are $g_{\beta_1}=2.6, 7.9, 13.2\,etc.$, and  $g_{\beta_2}$ is almost identical to $g_{\beta_1}$, as the difference between two $m^{*calc}_{\beta_1}$ and $m^{*calc}_{\beta_2}$ is small. 
Obtained $g_{\beta_1,0}=2.6$ differs a little from 2.7 obtained using the harmonic ratio method, which can be attributed to the anisotropy of $g$-factor.   

\subsection*{Anisotropic magnetoresitance}

\begin{figure}[h]
	\includegraphics[width=8cm]{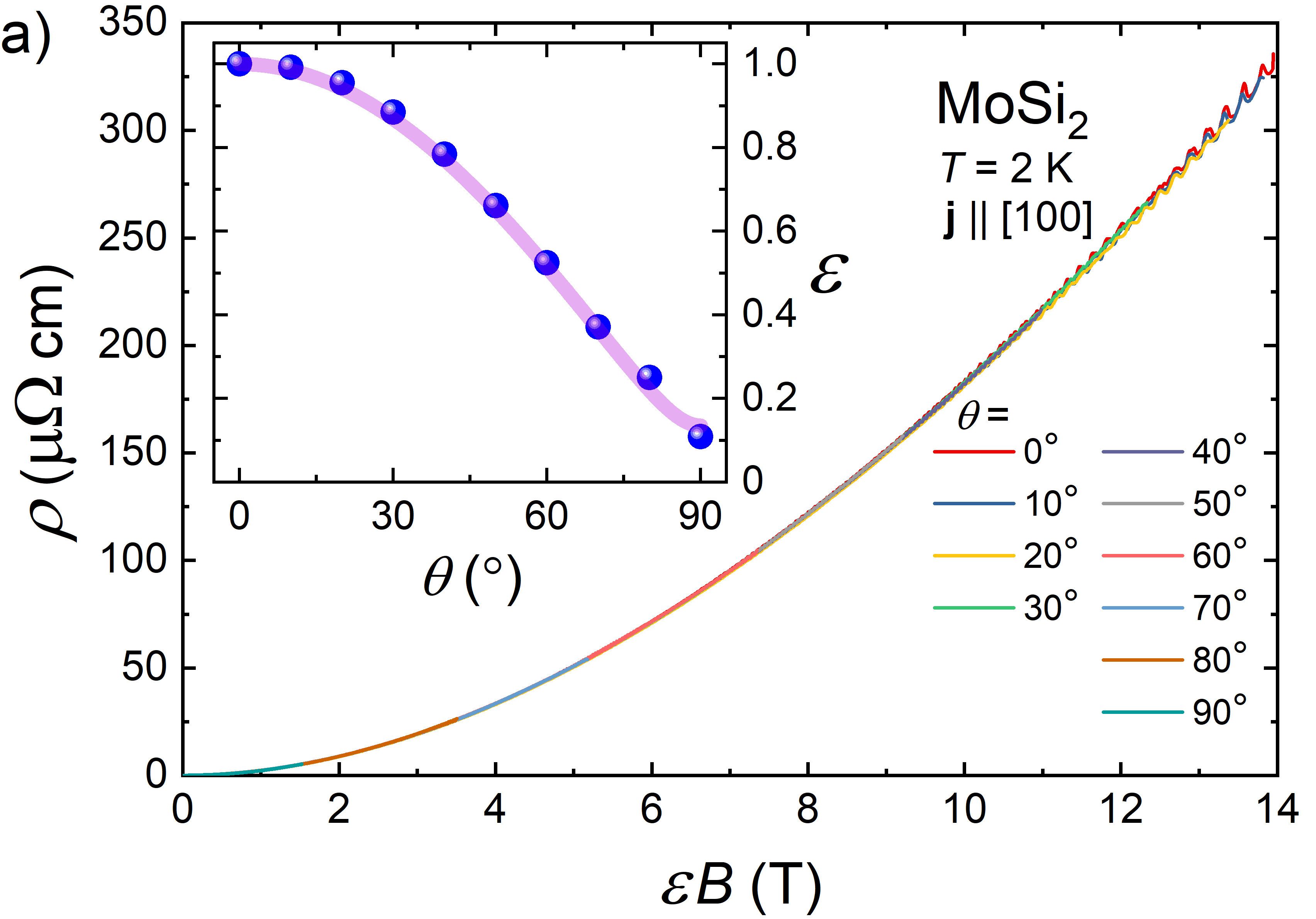}
	\includegraphics[width=8cm]{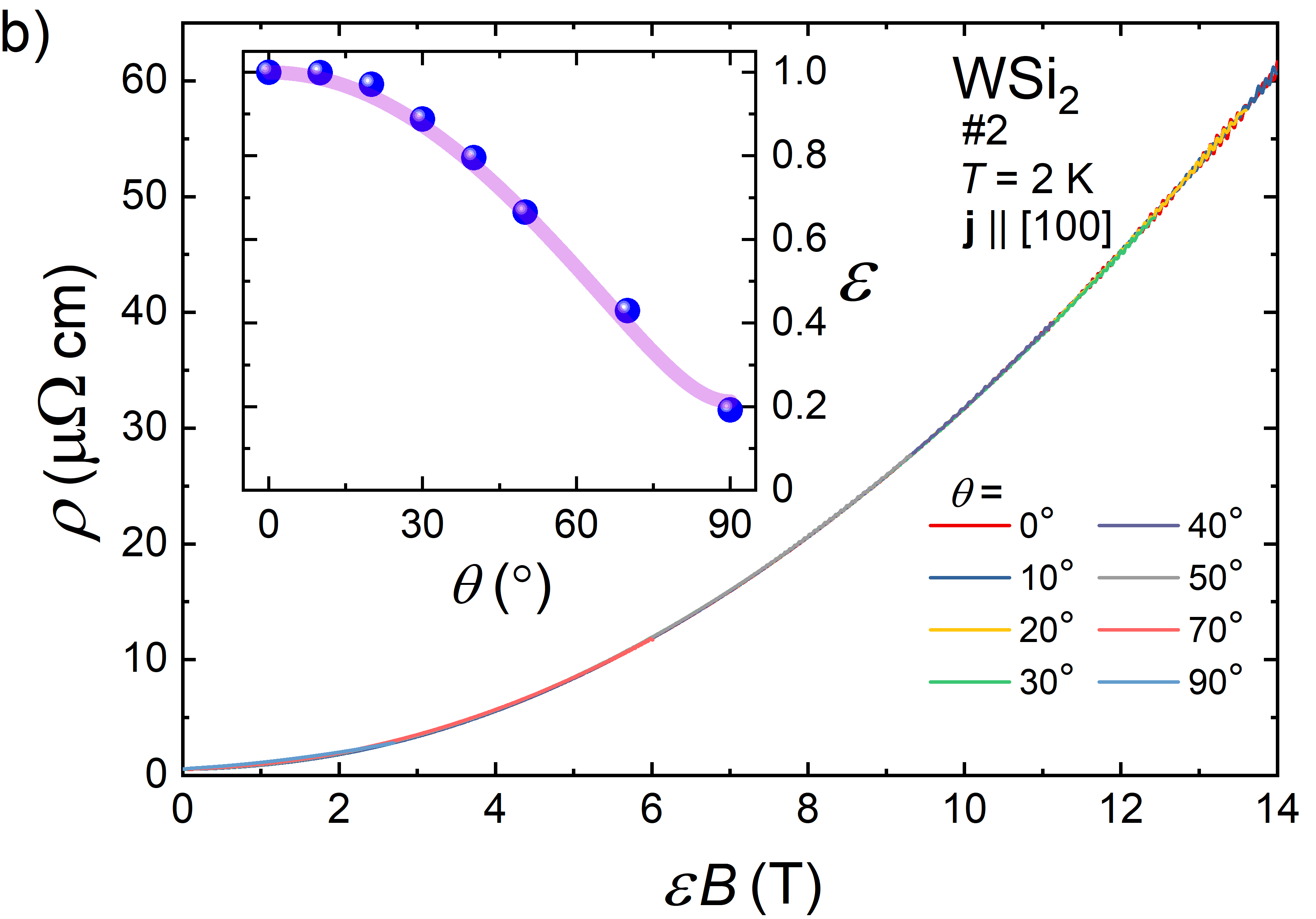}
	\caption{Electrical resistivity measured at $T = 2$\,K as a function of magnetic field scaled by factor $\epsilon_\theta$ for MoSi$_2$ (a) and WSi$_2$ (b), magnetic field was applied at different angles, $\theta$, to the current direction. Insets: $\epsilon_\theta$ as a function $\theta$ for MoSi$_2$ (a) and for WSi$_2$ (b). Violet solid lines correspond to the fits with equation~\ref{epsilon_theta_eq}.
		\label{rho(B)_epsilon_scaling}}
\end{figure}

A huge drop in the MR value of both materials, if compared to those measured in transverse configuration ($\textbf{B}\!\perp\!\textbf{j}$) to those recorded in longitudinal configuration ($\textbf{B}\!\parallel\!\textbf{j}$), can indicate that electrical resistivity is notably sensitive to the direction of the magnetic field application. 
This leads to large anisotropic magnetoresistance $({\rm {AMR}}=[\rho(90^{\circ})-\rho(0^{\circ})]/\rho(0^{\circ}))$ which is equal to $-95\%$ for WSi$_2$ and $-98\%$ for MoSi$_2$ at $T=2$\,K and in $B=14$\,T. 
The magnitudes of AMR are larger than those we reported previously for rare earth monoantimonides\cite{Pavlosiuk2016f,Pavlosiuk2017} and half-Heulser bismuthides.\cite{Pavlosiuk2019,Pavlosiuk2020,Pavlosiuk2021}
The origin of this huge AMR can be ascribed to the highly anisotropic Fermi surface of the studied compounds. 
Large AMR has been previously observed in materials with XMR, like bismuth,\cite{Zhu2011} graphite\cite{Soule1964} and WTe$_2$.\cite{Thoutam2015} 
The authors of the last work proposed a general scaling approach for materials with anisotropic Fermi surface. 
We successfully used this model in our recent works for describing the AMR of two NaCl-type crystal structure compounds YSb and LuSb (Refs~\onlinecite{Pavlosiuk2016f,Pavlosiuk2017}).
First, for each $\rho(B)$ curve, measured at particular $\theta$, the field values were scaled, so as all curves collapse on that corresponding to $\theta=0^\circ$ (see Fig.~\ref{rho(B)_epsilon_scaling}a,b). 
The so-obtained scaling factors ($\varepsilon$) plotted versus $\theta$ are shown in the insets Fig.~\ref{rho(B)_epsilon_scaling}a,b.  
According to the theory,\cite{Thoutam2015} $\varepsilon(\theta)$ depends on the parameter $\gamma$, which stands for the effective mass anisotropy:
\begin{equation}
	\varepsilon(\theta)=(\cos^2\theta+\gamma^{-2}\sin^2\theta)^{1/2}.
	\label{epsilon_theta_eq}
\end{equation}
Fittings of this equation to the experimental data are shown as violet solid lines in the insets to Fig.~\ref{rho(B)_epsilon_scaling}a,b and they give $\gamma=7.4$ and $\gamma=4.7$ for MoSi$_2$ and WSi$_2$, respectively. 
These value are larger than that reported for MoTe$_2$ in Ref.~\onlinecite{Chen2016a} as well as those we previously reported for LuSb and YSb in Refs~\onlinecite{Pavlosiuk2016f,Pavlosiuk2017}, respectively. 
For WSi$_2$ we obtained a value for $\gamma$ which is almost the same as $\gamma=4.762$ reported for WTe$_2$ (Ref.~\onlinecite{Thoutam2015}). 
The Fermi sheet anisotropy was also obtained based on the results of the SdH oscillations analysis as $k_{F,\beta}(\theta=90^\circ)/k_{F,\beta_2}(\theta=0^\circ)$, we got 4 for MoSi$_2$ and 3.3 for WSi$_2$.
These values are smaller than corresponding $\gamma$ values, probably due to the bold assumption that the hole-like pockets are cylinder-shaped.
It should be also noted that we do not take into account the anisotropy of the electron-like Fermi-pocket. 

\subsection*{Magnetostriction}

\begin{figure}[h]
	\includegraphics[width=8cm]{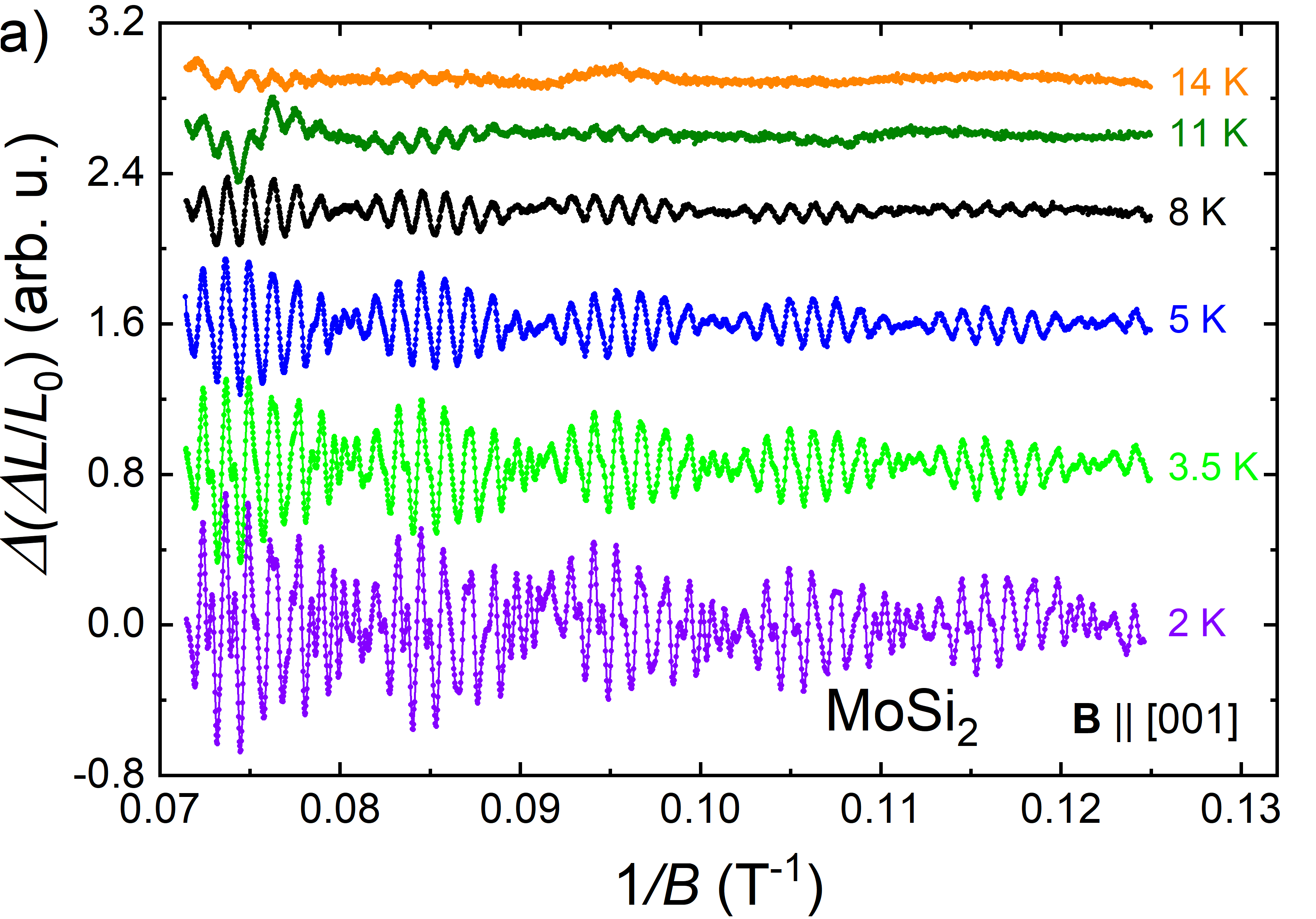}
	\includegraphics[width=8cm]{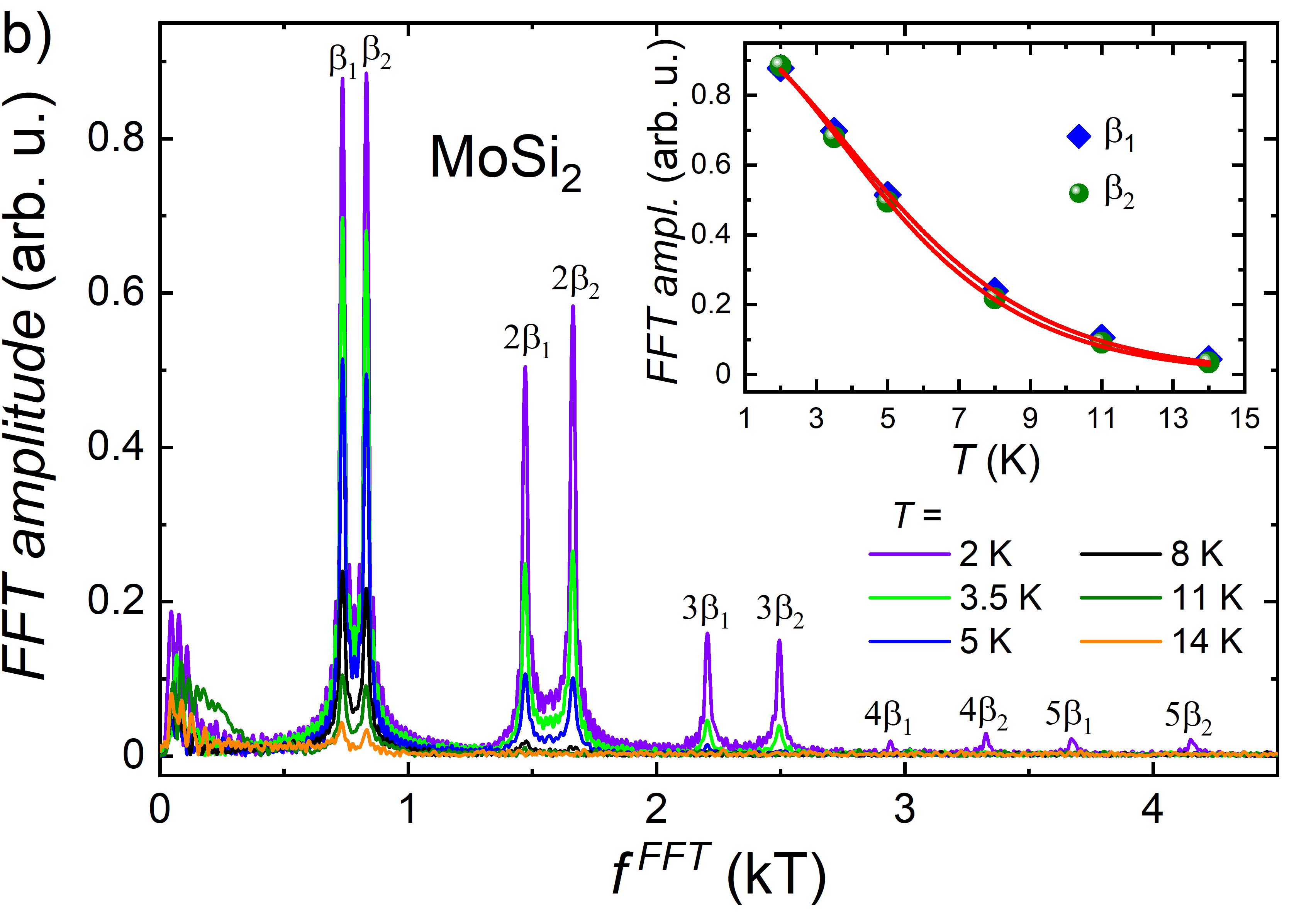}
	\caption{(a) Oscillating part of the magnetostriction of MoSi$_2$ as a function of inverted magnetic filed. (b) FFT spectra obtained from the analysis of data presented in (a). Inset: temperature dependence of the relative FFT amplitudes for both cross-sections, $\beta_1$ and $\beta_2$. Red solid lines correspond to the fit of temperature damping factor $R_T$ of LK equation (Eq.~\ref{full_LK_eq}) to the experimental data.    
		\label{Magnetostriction}}
\end{figure}

In our work we analyzed so far the SdH quantum oscillations of MoSi$_2$ and WSi$_2$ so far. 
In the literature, only dHvA oscillations in these compounds have been reported.\cite{Ruitenbeek1987,Matin2018,Mondal2020}  
These two quantum oscillations phenomena are most often described in literature but they belong to two different subclasses of quantum oscillations. 
The dHvA effect relates to thermodynamic properties, whereas SdH oscillations relates to non-equilibrium properties.\cite{Shoenberg1984}  
As the dHvA effect has been reported for both materials, we decided to look for relatively rarely reported oscillations of magnetostriction, another thermodynamic property. 

Magnetostriction in diamagnetic semimetals originates from the magnetic field induced changes of charge carrier concentration.\cite{Michenaud1982} 
We found that magnetic field induced length change, $\Delta L/L_0$, of MoSi$_2$ attains moderate values of the order $10^{-6}$ at $T=2$\,K and in magnetic field of 10\,T, which means that at these conditions carrier densities are not strongly affected by the magnetic field (see Supplementary Information). 
This value is similar to that observed for YAgSb$_2$ in Ref.~\onlinecite{Budko2008}, but is smaller than those found for TaAs in Ref.~\onlinecite{Cichorek2021} or LuAs in Ref.~\onlinecite{Juraszek2019}.
However, the most important feature is that at $B>4$\,T and at $T=2$\,K, the oscillating behavior of $\Delta L/L_0(B)$ is clearly observed.  
Fig.~\ref{Magnetostriction}a demonstrates the extracted quantum oscillations of magnetostriction of MoSi$_2$ at several temperatures for the range 2-14\,K. 
Their FFT analysis and plots of effective mass are presented in the main panel of Fig.~\ref{Magnetostriction}b and its inset, respectively.
The FFT spectra exhibit two fundamental frequencies $F^{FFT}_{{\beta_1}}=736$\,T and $F^{FFT}_{{\beta_2}}=831$\,T. 
Both of them are very close to those obtained from the SdH effect analysis. 
The small differences can be attributed mostly to tiny misalignments of the sample with respect to the magnetic field direction.  
At $T=2$\,K, we obtained rich spectrum of harmonic frequencies and even the 5th harmonics are clearly discernible. 
Unfortunately, we are not able to confirm that the combination frequency $F_{\beta_1-\beta_2}$ is also noticeable in our FFT spectra of magnetostriction oscillations due to some artifacts in the low-frequency range of our spectra. 
The calculated effective masses $m^*_{\beta_1}=\,0.285m_0$ and $m^*_{\beta_2}=\,0.3m_0$ are almost identical to those obtained from the SdH oscillations analysis (see Table~\ref{FFT_TABLE}).
As in the case of SdH oscillation, we also used the 3rd harmonic ratio method to estimate the lower limit of the $g$-factor of MoSi$_2$, based on the results of quantum oscillations of magnetostriction.\cite{Shoenberg1984} 
For both cross-sections, they are almost identical, $g_{0,\beta_1}\!\sim\! g_{0,\beta_2}\!\sim\!3$. All these values are slightly larger than those determined from the SdH oscillations analysis.

\subsection*{Conclusions} 

The present study was designed to investigate possible topologically non-trivial properties of electronic structure of two disilicides, WSi$_2$ and MoSi$_2$. 
First-principles quantum mechanical calculations are in favor of the existence  of tilted Dirac cones, located close to the Fermi level in both studied materials. 
This is the first report on the appearance of type-II Dirac states in these materials. 
Importantly, we found the substitution of tungsten by molybdenum leads to a shift in the Fermi level by around 200\,meV closer to the Dirac point compared to the pure WSi$_2$ compound. 
This finding suggests that in further research greater focus should be given to the chemical doping of both materials, because alloying could shift the Fermi level much closer to the Dirac point. 
The second major finding was that the presented results of analysis of angle-dependent Shubnikov-de Haas quantum oscillations are in a full agreement with the theoretically predicted electronic band structure.
The research has also shown that concentrations of electron-type carriers and hole-type carriers are almost perfectly balanced, indicating that charge compensation is responsible for the observed magnetotransport properties.

Analysis of the SdH oscillations of both compounds discloses the Shoenberg effect, surviving to relatively high temperatures of at least 25\,K and 12\,K for MoSi$_2$ and WSi$_2$, respectively. 
Additionally, for MoSi$_2$ the rare spin-zero effect is observed. 
In addition to SdH oscillations, we found the oscillating behavior of the magnetostriction of MoSi$_2$. 
The precise analysis of these quantum oscillations gave almost identical results to those obtained from the interpretation of SdH oscillations.
Finally, we discovered that extremely large anisotropic magnetoresistance recorded in both materials can be understood in the scope of the anisotropic Fermi surface.

\subsection*{Acknowledgements}

We are grateful to Ewa Bukowska for performing powder X-ray diffractography. 
O.P. was supported by the Foundation for Polish Science (FNP), program START 66.2020.
P.W.S. and J.-P.W. are supported in part by the Center for Spintronic Materials for Advance Information Technologies (SMART), one of seven centers of nCORE, a Semiconductor Research Corporation program.   
The band structure calculations were carried out simultaneously at the Interdisciplinary Centre for Mathematical and Computational Modelling (ICM) University of Warsaw under Grant No. GB76-4, and at the Wrocław Centre for Networking and Supercomputing under Grant No. 359. 
We thank Zach Cresswell for critical reading of the manuscript.

\newpage
\setcounter{figure}{0}
\renewcommand{\thefigure}{S\arabic{figure}}
\renewcommand{\thetable}{S\arabic{table}}
\renewcommand{\theequation}{S\arabic{equation}}
\bibliographystyle{naturemag}
\renewcommand{\citenumfont}[1]{S#1}
\renewcommand{\bibnumfmt}[1]{[S#1]}
\noindent\begin{large}{\bf Supplementary Information for} \end{large}\\\\
\begin{Large}{\bf Giant magnetoresistance, Fermi surface topology, Shoenberg effect and vanishing quantum oscillations in type-II Dirac semimetal candidates MoSi$_2$ and WSi$_2$}\end{Large}\\
	
\subsection*{Laue diffraction}

	\begin{figure}[h]
		\includegraphics[width=8cm]{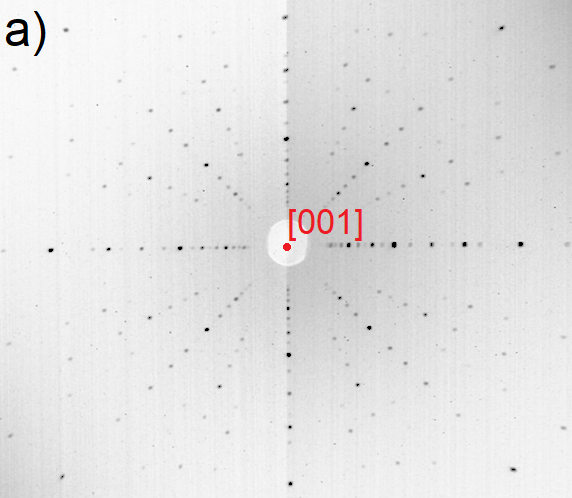}
		\includegraphics[width=8cm]{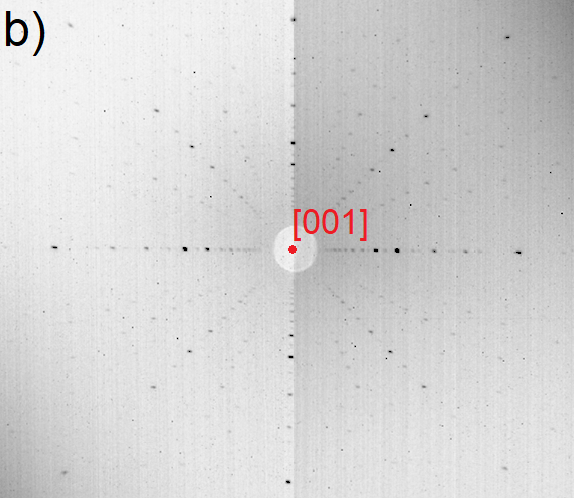}
		\caption{Laue patterns of MoSi$_2$ single crystal (a) and WSi$_2$ single crystal (b), with X-ray beam along [001] crystallographic direction.  
			\label{Lauegram}}
	\end{figure}
	\newpage
	\subsection*{Electrical resistivity}
	\begin{figure}[h]
		\includegraphics[width=8cm]{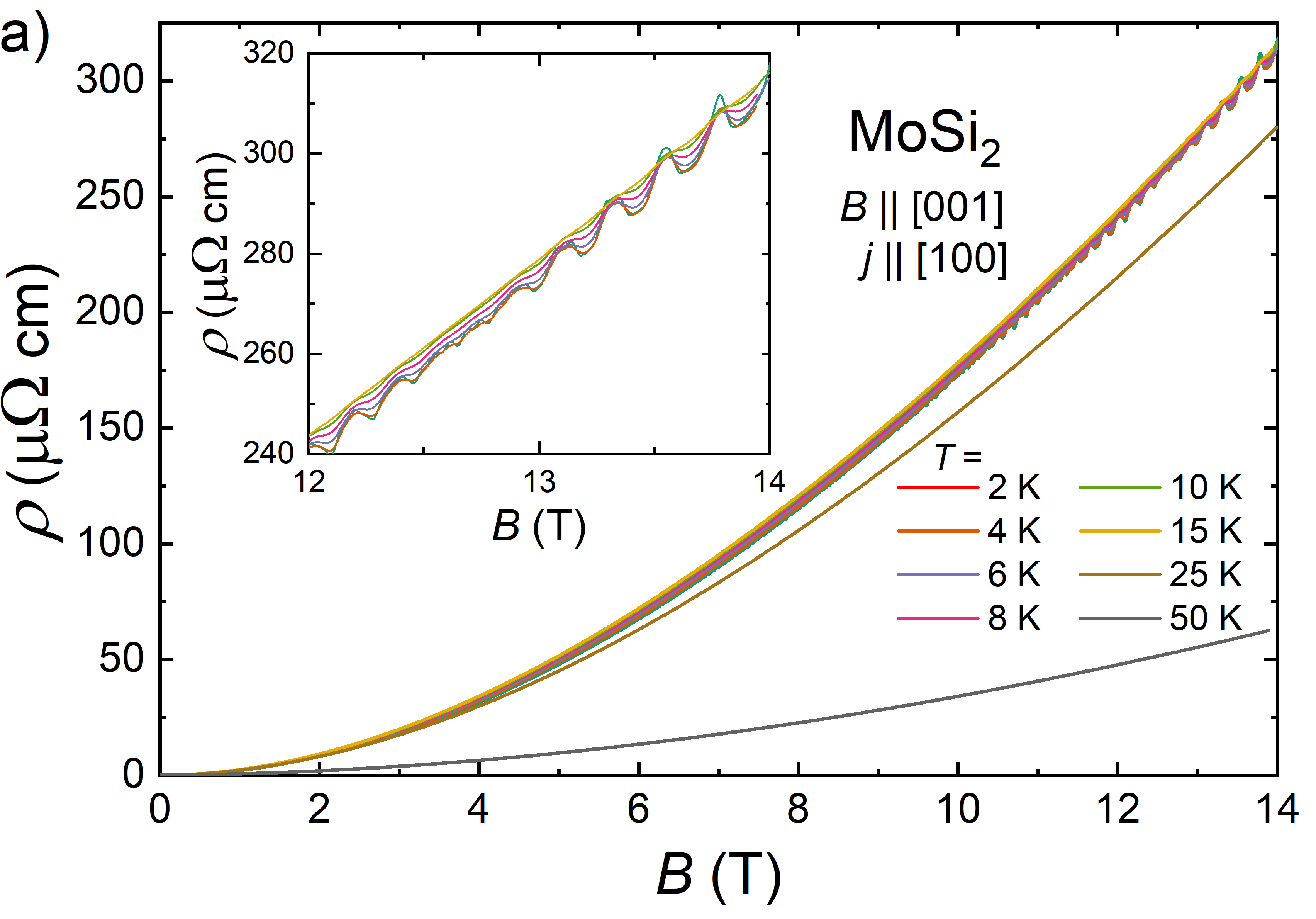}
		\includegraphics[width=8cm]{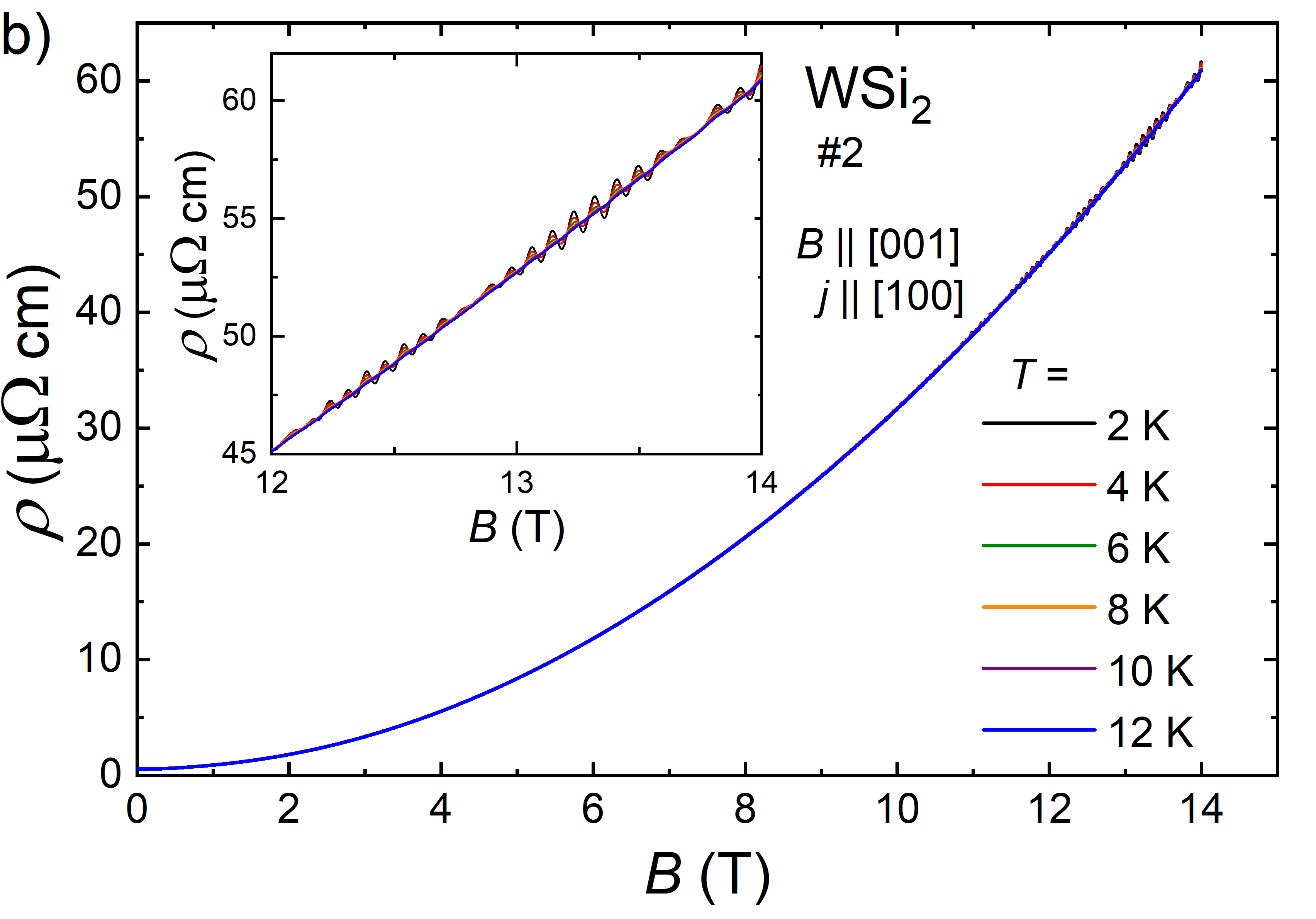}
		\caption{Electrical resistivity of MoSi$_2$ (a) and WSi$_2$ sample $\#2$ (b) as a function of magnetic field at several temperatures. Insets: Zoom of the data in the strong-magnetic-field range $12\,{\rm{T}}~\leq~B~\leq~14\,{\rm{T}}$; pronounced quantum oscillations are visible even in raw data.       
			\label{rho(T)}}
	\end{figure}
	
	\subsection*{FFT analysis for different intervals of magnetic field}
	\begin{figure}[h]
		\includegraphics[width=8cm]{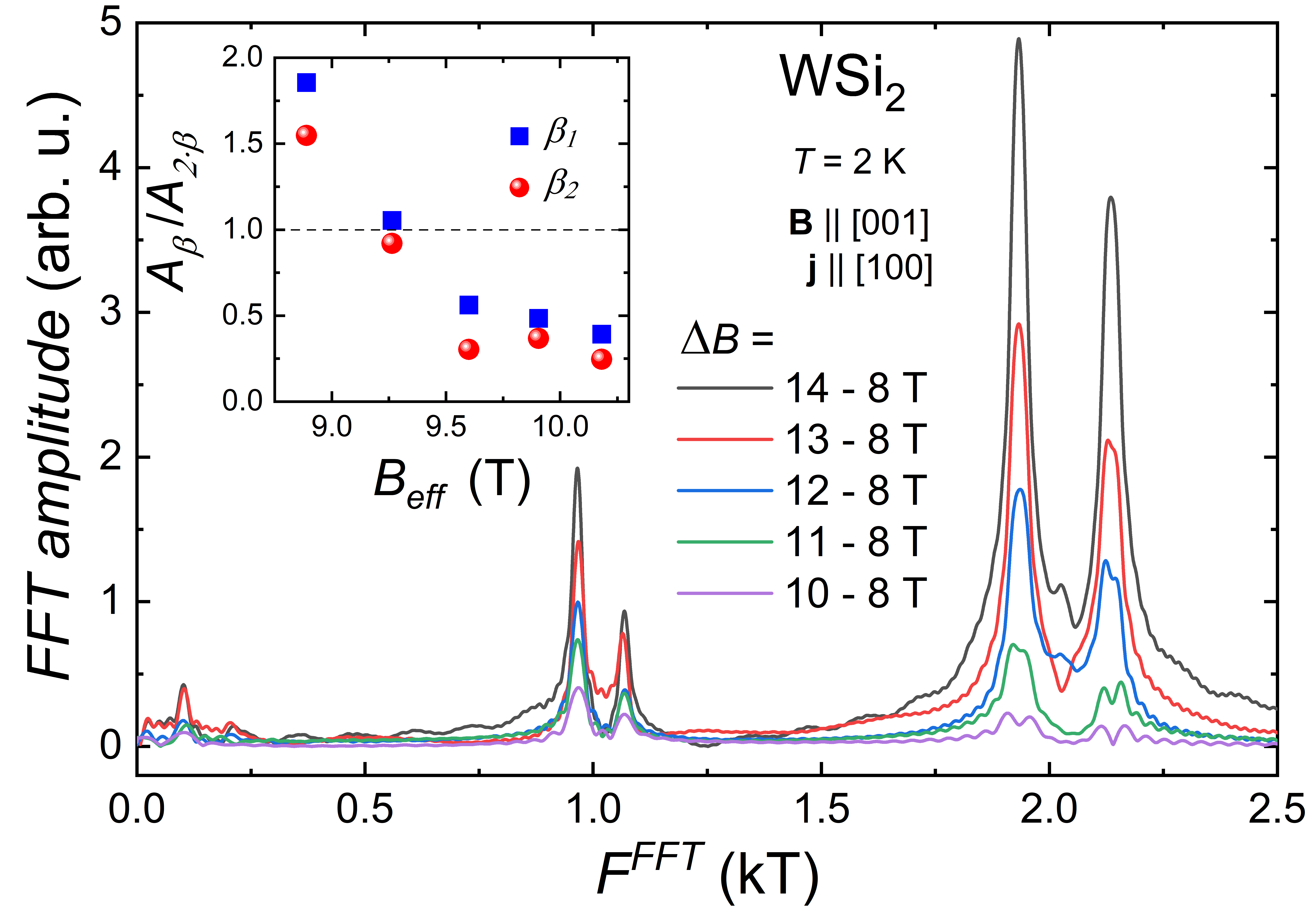}
		\caption{Fast Fourier transform spectra of quantum oscillations in WSi$_2$ at $T=2$\,K and in {\bf{B}}$\perp${\bf{j}} for several different intervals of magnetic field. Inset: ratio of the fundamental oscillation amplitude ($A_\beta$) to the amplitude of the second harmonic ($A_{2\beta}$) as a function of effective magnetic field ($B_{e\!f\!f}$). 
			\label{FFT_ranges}}
	\end{figure}

	We performed FFT analysis for several different magnetic fields' intervals (see Fig.~\ref{FFT_ranges}), starting from the range $8\,{\rm{T}}\!<\!B\!<\!14\,{\rm{T}}$ and gradually decreased the upper limit by 1\,T, achieving the final interval  $8\,{\rm{T}}\!<\!B\!<\!10\,{\rm{T}}$. 
	When the magnetic field decreases, the amplitude of the peaks also decreases, but peaks related to the second harmonic frequencies decrease rapidly. 
	To illustrate this process, we plot the ratio of the fundamental frequency amplitude ($A_\beta$) to the amplitude of the second harmonic ($A_{2\beta}$) as a function of effective magnetic field ($B_{e\!f\!f}$), defined as $B_{e\!f\!f}=2/(1/B_1+1/B_2)$, where $B_1$ and $B_2$ are the limits of the magnetic field interval of FFT. 
	For $B_{e\!f\!f}\!>\!9.5$\,T, amplitudes of the second harmonics are from 4 to 2 times larger than the amplitude of the fundamental frequency, but with further decrease of $B_{e\!f\!f}$ amplitude of the fundamental oscillation becomes larger than the amplitude of the second harmonic oscillation. 
	This behavior can be related to the decrease of the spin-reduction factor in the Lifshitz-Kosevich formula (Eq.~1 in the manuscript). 
	Assuming that effective mass does not significantly change in the studied magnetic field interval, responsibility for the decreasing of the spin-reduction factor lays on the Land\'e $g$-factor, which is attributed to the measure of the strength of the Zeeman effect.
	
	\subsection*{Hall effect}
	\begin{figure}[h]
		\includegraphics[width=8cm]{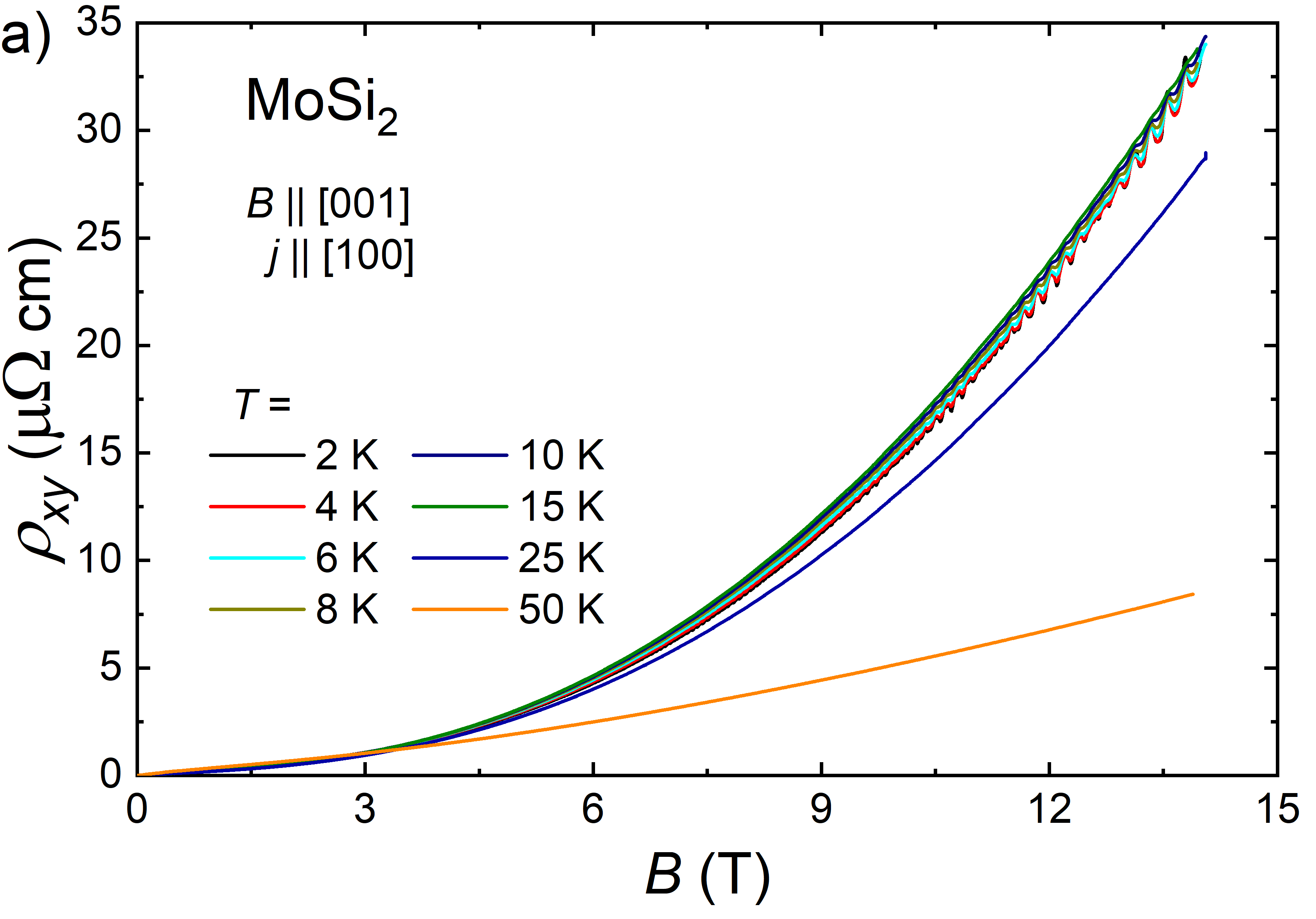}
		\includegraphics[width=8cm]{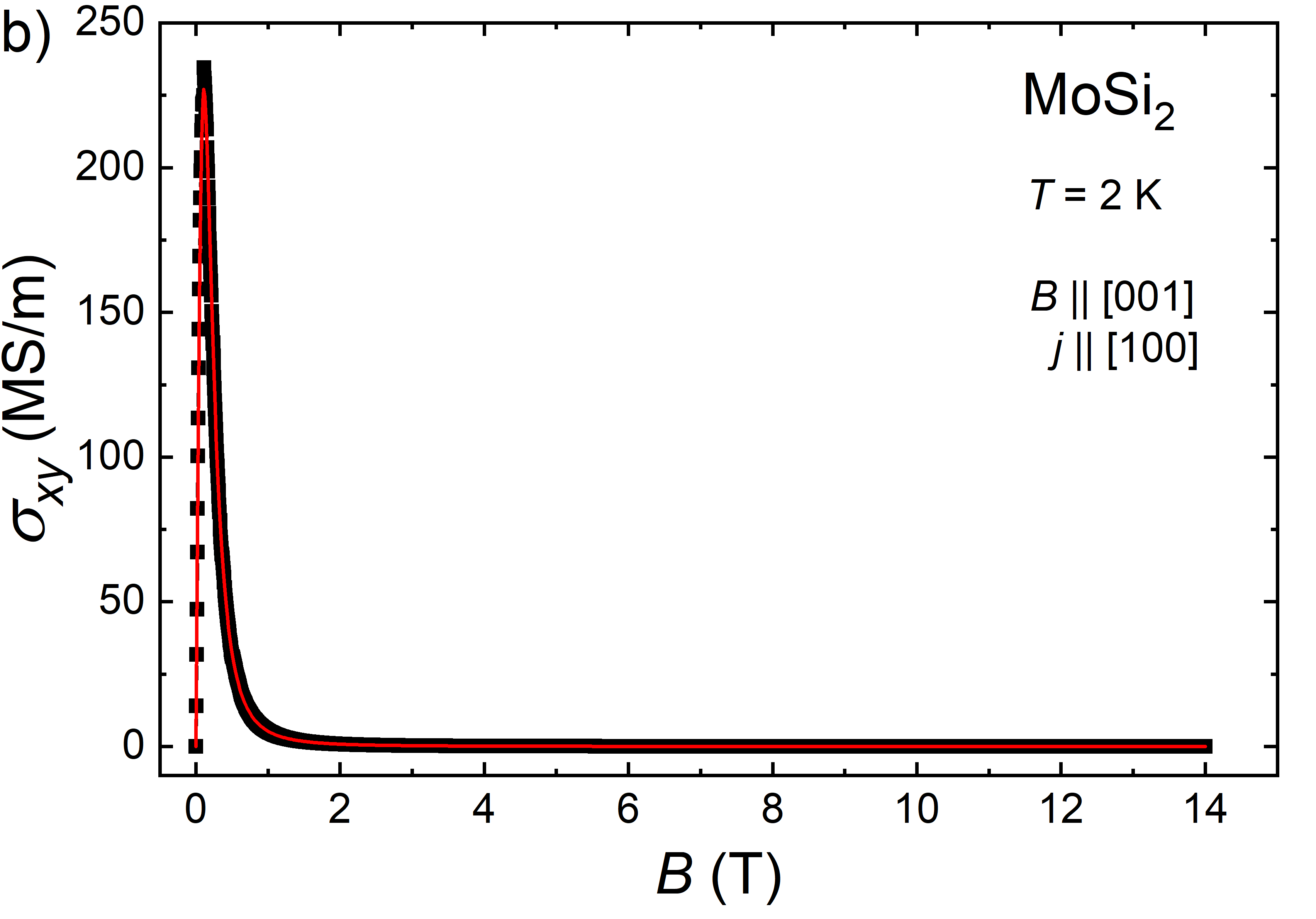}
		\caption{(a) Hall resistivity of MoSi$_2$ as a function of magnetic field taken at several temperatures from the range 2-50\,K. (b) Hall conductivity as a function of magnetic field at $T=2$\,K; the solid red line is a fit to Drude two-band model (Eq.~\ref{Eq_two-band_Drude})  
			\label{Hall}}
	\end{figure}

	The results of Hall effect measurements of MoSi$_2$ are in full agreement with both our theoretical calculations and quantum oscillations analysis results. 
	The Hall resistivity as a function of magnetic field, $\rho_{xy}(B)$, behaves in a manner typical for multi-carriers systems (see Fig.~\ref{Hall}a). 
	In contrast to Ref.~\cite{Matin2018_SM}, in our sample carriers of a hole-type dominate the transport properties (positive values of $\rho_{xy}$ for $B>0$), which means that position of the Fermi level of both samples critically depends on the degree of crystal structure imperfection.
	To characterize quantitatively the electronic transport in our single crystal, we applied the two-band Drude model~\cite{Hurd1972_SM}.
	Within this approach, Hall conductivity ($\sigma_{xy}=-\rho_{xy}/(\rho_{xx}^2+\rho_{xy}^2)$) of a semimetal can be described by the equation:
	\begin{equation}
		\sigma_{xy}=eB\bigg(\frac{n_h\mu_h^2}{1+(\mu_hB)^2}+\frac{n_e\mu_e^2}{1+(\mu_eB)^2}\bigg),
		\label{Eq_two-band_Drude}
	\end{equation} 
	where $e$ is the elementary charge; $n_h$ and $n_e$ stand for the carrier concentrations of holes and electrons, $\mu_h$ and $\mu_e$ represent their mobilities. 
	The result of fitting Eq.~\ref{Eq_two-band_Drude} to the experimental data is shown by the red solid line in Fig.~\ref{Hall}b. 
	The analysis yielded the following parameters: $n_h=9.41\!\times\!10^{20}\,\rm{cm^{-3}}$, $n_e=9.38\!\times\!10^{20}\,\rm{cm^{-3}}$, $\mu_h=6.42\,\!\times\!10^{4}\rm{cm^2V^{-1}s^{-1}}$ and $\mu_e=4.08\,\!\times\!10^{4}\rm{cm^2V^{-1}s^{-1}}$. 
	The obtained carrier concentrations are in a favour of perfect carrier compensation in this system ($n_e/n_h=0.997$). 
	The values of $n_h$ and $n_e$ are of the same order of magnitude as those reported recently in Ref.~\cite{Matin2018_SM}. However, the estimated mobility of electrons is one order of magnitude smaller. 
	In Ref.~\cite{Matin2018_SM}, the authors claimed that magnetic field changes the ratio of different types of carriers in their sample, however, we did not observe such a behaviour in our samples.
	Furthermore, our results of magnetostriction measurements (see below) indicate that the carrier concentrations are hardly sensitive to magnetic field.          
	
	\subsection*{Magnetostriction}

	\begin{figure}[h]
		\includegraphics[width=8cm]{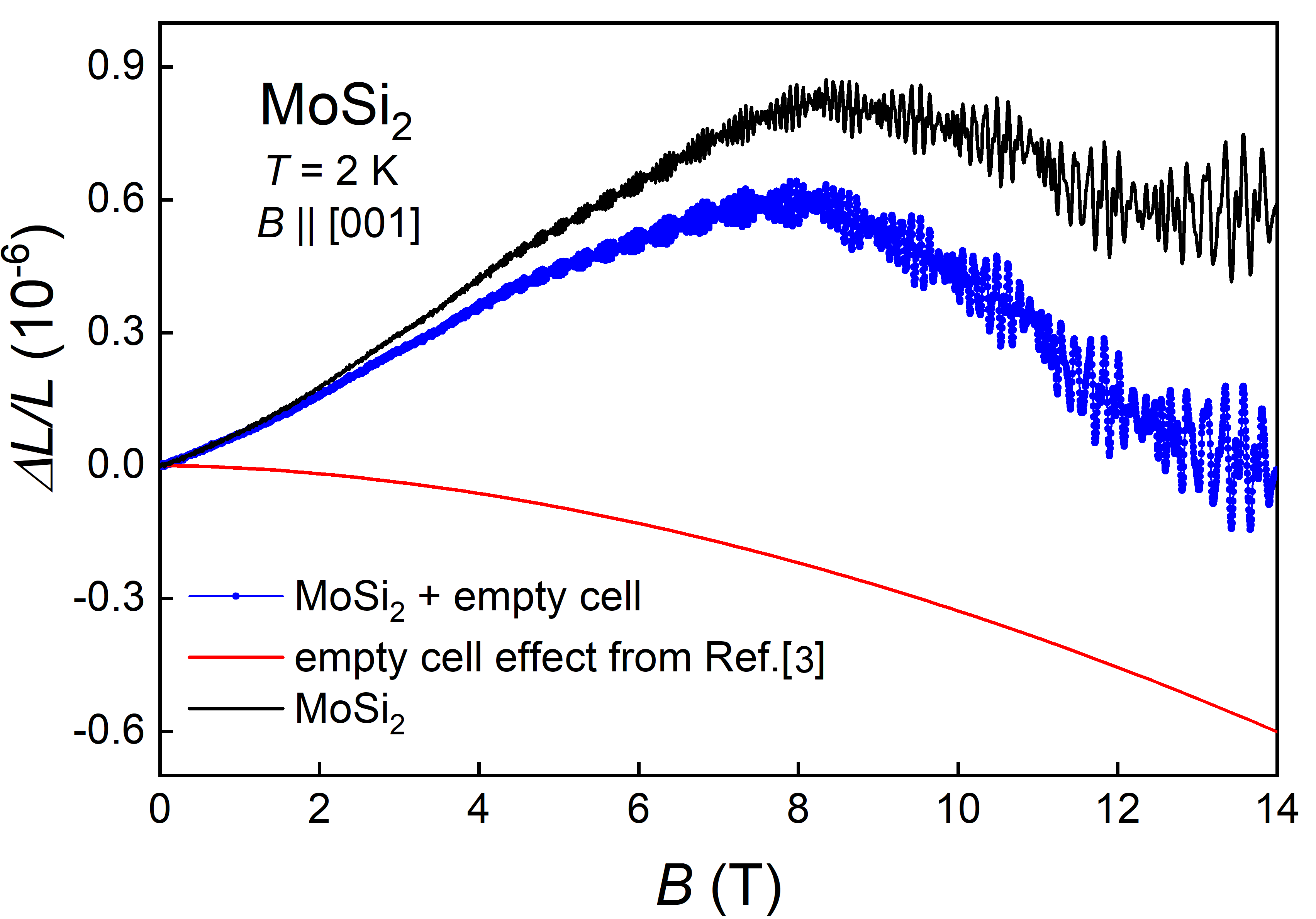}
		\caption{Magnetic field dependent relative length change ($\Delta L/L_0$) of the MoSi$_2$ sample obtained at $T=2$\,K with magnetic field parallel to the [001] crystallographic direction (solid black curve). The solid red curve stands for the empty cell effect described as $\Delta L(B)=-5.2\!\times\!10^{-9}B^{1.8}$, taken from Ref.~\cite{Kuchler2012_SM}. 
			The blue curve corresponds to the sum of both contributions, empty cell and MoSi$_2$.   
			\label{Magnetostriction}}
	\end{figure}

	Figure~\ref{Magnetostriction} demonstrates the procedure of the extraction of the magnetic field dependent relative length change of MoSi$_2$ sample at $T=2$\,K. 
	The smooth background which is described by the relation $\Delta L(B)=-5.2\!\times\!10^{-9}B^{1.8}$ (see the red solid curve in Fig.~\ref{Magnetostriction}) and corresponds to the empty cell effect was subtracted from the total signal measured (the blue curve in Fig.~\ref{Magnetostriction}). 
	The so-obtained pure contribution of MoSi$_2$ is shown as the  black solid line.
	$\Delta L/L_0(B)$ increases with increasing magnetic field up to $B\sim\!8$\,T and then decreases. 
	The magnitude of $\Delta L/L_0$ of MoSi$_2$ is relatively small being only $9\times10^7$ at the maximum, which is one order of magnitude smaller than the values reported for e.g. Bi~\cite{Michenaud1982_SM} or LuAs~\cite{Juraszek2019_SM}, but comparable to these measured for YAgSb$_2$~\cite{Budko2008_SM}.  
	This means that magnetic field changes the carrier density in MoSi$_2$ in a fairly moderate manner. 
	Assuming that the magnetostriction of a diamagnetic semimetal can be described as $\Delta L/L_0(B)=c\Delta n(B)$, where a material-dependent coefficient, $c$, equals to $1.5\times10^{-24}\,\rm{cm^3}$ (Ref.~\cite{Michenaud1981_SM}), we found that in $B=8$\,T, $\Delta n_h/n_h\approx\Delta n_e/n_e\approx0.06$. 
	This rules out the effect of magnetic-field driven XMR, anticipated for MoSi$_2$ in Ref.~\cite{Matin2018_SM}.

\newpage	
\noindent{\bf Supplemental references}

\end{document}